\documentclass[twocolumn,showpacs,superscriptaddress,amsmath,amssymb,prl,10pt,showkeys,aps,longbibliography]{revtex4-1}
\usepackage{graphicx}
\usepackage{dcolumn}
\usepackage{bm}
\usepackage{hyperref}
\usepackage{amsthm}
\usepackage{wasysym}
\usepackage{kantlipsum}
\usepackage{booktabs}
\newcolumntype{C}{>{$}c<{$}}
\AtBeginDocument{
\heavyrulewidth=.08em
\lightrulewidth=.05em
\cmidrulewidth=.03em
\belowrulesep=.65ex
\belowbottomsep=0pt
\aboverulesep=.4ex
\abovetopsep=0pt
\cmidrulesep=\doublerulesep
\cmidrulekern=.5em
\defaultaddspace=.5em
}

\usepackage{siunitx}
\usepackage{multirow}

\usepackage[normalem]{ulem}
\usepackage[usenames,dvipsnames]{color}

\makeatletter
\newcommand*{\addSI}{%
  \close@column@grid
  \cleardoublepage
  \twocolumngrid
}
\makeatother

\def\eqref#1{(\ref{#1})}
\def\angstrom{{\mbox{\AA}}}
\newcommand{\un}[1]{\,\mathrm{#1}}

\newtheorem*{theorem*}{Theorem}
\newtheorem*{corollary*}{Corollary}

\definecolor{tangerine}{rgb}{0.944,0.522,0}
\definecolor{verde}{rgb}{0.,0.6,0}
\definecolor{rosso}{rgb}{0.9,0.0,0.2}
\definecolor{magenta}{rgb}{0.9,0.2,0.9}

\newcommand{\editor}[2]{%
  \expandafter\newcommand\csname #1note\endcsname[1]{%
    \textcolor{#2}{(\textbf{#1:} ##1)}}%
  \expandafter\newcommand\csname #1\endcsname[1]{%
    \textcolor{#2}{##1}}%
  \expandafter\newcommand\csname #1cancel\endcsname[1]{%
    \textcolor{#2}{\sout{##1}}}%
  \expandafter\newcommand\csname #1change\endcsname[2]{%
    \textcolor{#2}{\sout{##1} ##2}}%
  \newenvironment{#1text}{\color{#2}}{\color{black}}
}

\editor{SB}{tangerine}
\editor{LS}{verde}
\editor{FG}{rosso}

\begin{document}

\title{Heat {and charge} transport in H$_2$O at ice-giant conditions \\ from ab initio {molecular dynamics} simulations}

\author{Federico Grasselli}
\altaffiliation{Present affiliation: COSMO -- Laboratory of Computational Science and Modelling, IMX, \'Ecole Polytechnique F\'ed\'erale de Lausanne, 1015 Lausanne, Switzerland}
\affiliation{SISSA -- Scuola Internazionale Superiore di Studi Avanzati, Trieste, Italy}
\author{Lars Stixrude}
\affiliation{Department of Earth, Planetary, and Space Sciences, University of California Los Angeles, USA}
\author{Stefano Baroni}\email{baroni@sissa.it}
\affiliation{SISSA -- Scuola Internazionale Superiore di Studi Avanzati, Trieste, Italy}
\affiliation{CNR -- Istituto Officina dei Materiali, SISSA, 34136 Trieste}


\begin{abstract}
  {The impact of the inner structure and thermal history of planets on their observable features, such as luminosity or magnetic field, crucially depends on the poorly known heat and charge transport properties of their internal layers. The thermal and electric conductivities} of different phases of water (liquid, solid, and super-ionic) occurring in the interior of ice giant planets{, such as Uranus or Neptune,} are evaluated from equilibrium \emph{ab initio} molecular dynamics, leveraging recent progresses in the theory and data analysis of transport in extended systems. {The implications of our findings on the evolution models of the ice giants are briefly discussed}. 

\end{abstract}

\maketitle

\section{Introduction}

Hydrogen and oxygen are two of the three most abundant elements in the universe, helium being the second. As a consequence, H$_2$O is thought to be a major constituent of celestial bodies formed far enough from their host star for it to condense \cite{Lodders2003}. Many moons of the outer solar system, such as Ganymede, Europa, and Enceledus, have rigid icy shells and interior water oceans, which are key for understanding the observed surface mass flux \cite{Nimmo2007} and the generation of magnetic fields \cite{Kivelson1996}. The ice giants, Uranus and Neptune, are thought to be composed primarily of H$_2$O \cite{Nettelmann2013}: throughout most of their interior, the large pressure and temperature (\emph{e.g.} 240 GPa and 5000 K at half the radius of Uranus) favor a super-ionic (SI) phase, where oxygen ions are arranged in a crystalline lattice and protons diffuse freely like in a fluid \cite{Cavazzoni1999,Millot2018}. Partially-dissociated, liquid (PDL) water may instead be confined to the outermost third of the interior, where the magnetic field is generated \cite{Stanley2004}. Outside the solar system, the observed characteristics of many exoplanets are also consistent with with water-rich interiors \cite{Zeng2019}.

{Our knowledge of the interior of planets other than Earth mostly relies on the observation of their magnetic fields and surface properties, which are affected by the inner structure through the transport of energy, mass, and charge across intermediate layers. In the case of Uranus, for instance, it has long been recognized that its remarkably small luminosity \cite{Fortney2010} can be explained by non-adiabatic models of the interior \cite{Hubbard:1995,Nettelmann2013}, featuring thermal boundary layers whose transport properties are poorly known. Likewise, any model aiming to explain the anomalous multipolar and non-axisymmetry magnetic fields of Uranus and Neptune requires the knowledge of the electric conductivity of the various phases of water occurring in their interior \cite{Helled2020}. More generally, a detailed knowledge of the transport properties of different phases of H$_2$O occurring at high-pT conditions is key to any quantitative evolutionary model of water-rich celestial bodies.}
In spite of the steady progress in diamond-anvil-cell and shock-wave technologies, the experimental investigation of transport properties of materials at planetary conditions is still challenging. In the specific case of H$_2$O, the electrical conductivity is only known with large uncertainties along the Hugoniot curve on a limited portion of the pT diagram, and nothing is known about the heat conductivity \cite{Mitchell1982,Yakushev2000,Chau2001,Millot2018}.

Computer simulations may be our only handle on the properties of matter at physical conditions that cannot be achieved in the laboratory. In the case of water, they have allowed us to discover new phases \cite{Cavazzoni1999} and to predict their properties at extreme pT conditions \cite{Millot2018,Rozsa2018} over an ever broader portion of its phase diagram \cite{Sun2015}. The diverse local chemical environments that characterize the different relevant phases of water make classical force fields unfit for an accurate simulation of their properties, and call for a full quantum-mechanical, \emph{ab initio} (AI), treatment of the chemical bond. Some transport properties of water at high pT conditions, such as ionic (H and O) diffusivities and the electrical conductivity have indeed been estimated using AI molecular dynamics (AIMD) simulations \cite{French2011} and the Green-Kubo (GK) theory of linear response \cite{Green1952,*Green1954,Kubo1957b,*Kubo1957a}. However, it has long and widely been argued that quantum-mechanical simulation methods could not be combined with the GK theory, because the latter is based on a microscopic representation of the energy (current) density, which is evidently ill-defined at a quantum mechanical level \cite{Stackhouse2010b}. The soundness of this objection, which would actually apply to a classical representation of the interatomic forces as well, was recently refuted for good by the introduction of a \emph{gauge invariance} principle for transport coefficients \cite{Marcolongo2016,Ercole2016,Baroni2018}. In a nutshell, gauge invariance means that transport coefficients do not depend on the details of the microscopic representation of the conserved quantity being transported, as long as this representation sums to the correct value in the thermodynamic limit and its space correlations are short-ranged. This remarkable finding implies that \emph{any} (good, in the above sense) local representation of the energy leads to the same heat conductivity, thus paving the way to a fully \emph{ab initio} treatment of heat transport \cite{Marcolongo2016}, which was recently generalized to multi-component systems \cite{Bertossa2019}.

In this work we leverage these recent theoretical advances to estimate the thermal conductivity and other transport coefficients of stoichiometric H$_2$O in the pT conditions to be found on ice giant planets, like Uranus and Neptune, from equilibrium AIMD simulations, exploring its solid, PDL, and SI phases.

\section{Results}

\subsection{Theory}
Transport in macroscopic media is governed by the dynamics of hydrodynamic variables, \emph{i.e.} by the long-wave\-length components of the (current) densities of conserved extensive quantities \cite{Kadanoff1963,*Foster1975,Baroni2018}. For short, we will dub such densities \emph{conserved densities}, the corresponding currents \emph{conserved currents}, while the macroscopic averages of the latter will be called \emph{conserved fluxes}. The GK theory of linear response \cite{Green1952,Green1954,Kubo1957a,Kubo1957b} states that transport coefficients (\emph{i.e.} conductivities) are integrals of the various flux time autocorrelation functions, which, according to the Wiener-Khintchine theorem \cite{Wiener1930, Khintchine1934}, are the zero-frequency values of the corresponding power spectra. An important concept in the theory of transport is that of \emph{diffusive flux}: we say that a flux is \emph{diffusive} if its power spectrum does not vanish \cite{Baroni2018,Bertossa2019} at zero frequency. Gauge invariance states that two different representations (``\emph{gauges}'') of a same conserved density that differ by the divergence of a bounded vector field are equivalent in that they give rise to macroscopic fluxes whose difference is non-diffusive, thus resulting in the same conductivity \cite{Marcolongo2016,Ercole2016}.

When addressing heat transport, the relevant conserved quantities are the energy and the numbers of particles (or masses) of \textit{each} atomic species. Since the total-mass flux itself (\emph{i.e.} the total momentum) is a constant of motion, for a $P$-species system  the number of \textit{independent} conserved fluxes is equal to $P$ (energy, plus $P-1$ partial masses). Further constraints may reduce the number of relevant conserved fluxes. For instance, in solids, such as ordinary ice, atoms do not diffuse and there cannot be any macroscopic mass flow: energy flux is therefore the only relevant one. In molecular liquids, such as ordinary water, the partial mass fluxes of each atomic species are non-diffusive if the molecules do not dissociate. This is so because the integral of the difference between the individual total momenta of different atomic species is bound by the finite variation of the molecular bond lengths \cite{Marcolongo2016,Baroni2018}: also in this case, therefore, energy is the only relevant conserved quantity. On the contrary, the PDL and SI phases of water are truly multi-component systems, because the momentum of at least one atomic component is neither conserved nor is its integral bound by any molecular constraints.

Heat transport in multi-component systems has long been the subject of theoretical misconceptions and/or considered to be numerically intractable. For instance, the thermal conductivity is sometimes computed as the {GK} integral of the energy flux, $\mathbf{J}_E$:
 $   \kappa \propto \int_0^\infty \langle \mathbf{J}_E(t) \mathbf{J}_E(0) \rangle dt. $
This simplistic approach is manifestly wrong, as the resulting conductivity would depend on the arbitrary choice of the atomic formation energies. To see why this is so, let us consider the classical expression of the energy flux \cite{Irving1950,Baroni2018}:
$
   \mathbf{J}_E = \frac{1}{\Omega} \left[ \sum_{n=1}^N \mathbf{V}_n \epsilon_n
   + \sum_{n,m} (\mathbf{R}_n - \mathbf{R}_{m}) \mathbf{F}_{nm}\cdot \mathbf{V}_{n} \right],
$
where $\Omega$ is the system's volume, $\mathbf{R}_n$, $\mathbf{V}_n$, and $\epsilon_n$ are the atomic positions, velocities, and energies, respectively, and $\mathbf{F}_{nm} = -\partial \epsilon_{m} / \partial \mathbf{R}_n$ are inter-atomic forces.
The heat conductivity cannot evidently depend on the arbitrary zero of the atomic energies. For instance, in \emph{ab initio} calculations these energies differ in a pseudo-potential or in an all-electron scheme, whereas transport coefficients should not.
A better choice would be to compute the heat conductivity from the {GK} integral of the \emph{heat} flux, defined as $\mathbf{J}_q = \mathbf{J}_E - \frac{1}{\Omega}\sum_{S=1}^P h_S N_S \overline{\mathbf{V}}_S$, where {$\overline{\mathbf{V}}_S$ is the center-of-mass velocity and $h_S$} the partial enthalpy of the $S$-th atomic species \cite{DeGroot1984}. This approach has the advantage that $\mathbf{J}_q$ is no longer sensitive to a rigid shift in the atomic energies; it is still an approximation, though, as it neglects the coupling between energy and mass flow (Soret effect) in the calculation of $\kappa$. Even if, for several systems, it has been argued that the error in doing so is small \cite{DeGroot1984}, this argument cannot be taken for granted \textit{a priori} for any generic system. Furthermore, the calculation of partial enthalpies is rather involved \cite{Debenedetti1987,Vogelsang1987,Sindzingre1989}, and often the subject itself of crude approximations.

A rigorous methodology to deal with multi-component systems is provided by a combination of Onsager's phenomenological approach \cite{Onsager1931a,*Onsager1931b} and the GK theory of linear response \cite{Green1952,Green1954,Kubo1957a,Kubo1957b}. In this approach the interactions among different conserved fluxes are explicitly accounted for {by Onsager's phenomenological relations}:
\begin{equation}
  J_i = \sum_j \Lambda_{ij} f_j, \label{eq:Onsager}
\end{equation}
where $J$ is a generic conserved flux, $f$ a  thermodynamic affinity, \emph{i.e.} the average gradient of the intensive variable conjugate to a conserved quantity, {$\Lambda$ is the matrix of Onsager's phenomenological coefficients,} and the suffixes enumerate in principle both different conserved quantities and the Cartesian components of their fluxes/affinities. In practice, in the following we will dispose of Cartesian components, and only enumerate different conserved fluxes/affinities, given that we will only be concerned with isotropic or cubic systems. Within the GK theory, and the $\Lambda$ coefficients are expressed as integrals of the time correlation functions of the relevant fluxes:
\begin{equation}
    \Lambda_{ij} = \frac{\Omega}{k_B} \int_0^\infty \left\langle \mathcal{J}_i(t) \mathcal{J}_j(0) \right \rangle dt, \label{eq:L_def}
\end{equation}
where $\mathcal{J}_i(t)$ is the time series of the $i$-th flux{, $k_B$ is the Boltzmann constant, and $\langle\cdot\rangle$ indicates an equilibrium average}. From now on, calligraphic fonts indicate samples of stochastic processes. The thermal conductivity is defined as the ratio between the energy flux and the temperature gradient, \emph{when all the other conserved fluxes vanish}. In a two-component system this condition leads to the following expression for the heat conductivity:
\begin{equation}
   \kappa = \frac{1}{T^2}\left[ \Lambda_{EE} - \frac{\left| \Lambda_{EM}\right|^2}{\Lambda_{MM}} \right], \label{eq:multi_kappa}
\end{equation}
where the $M$ suffix indicates the mass flux of one of the two components. The expression in square brackets is the inverse of the $EE$ matrix element of the inverse of the $2\times 2$ matrix of the Onsager coefficients. In the general, multivariate, case, the heat conductivity is proportional to the \emph{Schur complement of the mass block} in $\Lambda$. In Ref. \onlinecite{Bertossa2019} we have shown that this expression for the heat conductivity is invariant under the addition of an arbitrary linear combination of conserved fluxes (such as mass or adiabatic electronic charge) to the energy flux, and we named this further remarkable property of transport coefficients \emph{convective invariance}.

Eq. \eqref{eq:multi_kappa} shows that this procedure is numerically ill-conditioned, because the estimator of the integral in Eq. \eqref{eq:L_def} becomes a random walk as a function of the upper limit of integration, as soon as the integrand has exhausted all its weight, thus making the expression in Eq. \eqref{eq:multi_kappa} singular whenever the estimator of the denominator vanishes \cite{Galamba2007,Ohtori2009b,Salanne2011,Bonella2017}. A solution to this problem is provided by \emph{multivariate cepstral analysis} \cite{Bertossa2019}, briefly sketched below. According to the Wiener-Khintchine theorem \cite{Wiener1930,Khintchine1934}, the Onsager coefficients in Eq. \eqref{eq:L_def} are proportional to the zero-frequency values of the flux cross power spectral density, $S_{ij}(\omega)=\int_{-\infty}^\infty \left\langle \mathcal{J}_i(t) \mathcal{J}_j(0) \right \rangle \mathrm{e}^{i\omega t} dt$:
\begin{align}
  \Lambda_{ij} & =\frac{\Omega}{2k_B}S_{ij}(\omega=0) \\
  S_{ij}(\omega) &= \lim_{\tau\to\infty} \langle \mathcal{S}^\tau_{ij}(\omega) \rangle \label{eq:GKM-1}\\
  \mathcal{S}^\tau_{ij}(\omega) &= \frac{1}{\tau} \tilde {\mathcal{J}}^\tau_i(\omega)^* \cdot \tilde {\mathcal{J}}^\tau_j(\omega) \label{eq:GKM-2}\\
  {{\tilde{\mathcal{J}}}}_j^\tau(\omega)&=\int_0^\tau \mathcal{J}_j(t) \mathrm{e}^{i\omega t}dt. \label{eq:GKM-3}
\end{align}
The continuity and smoothness of the power spectrum at low frequency can be leveraged to systematically reduce the noise affecting the estimator of its zero-frequency value, as explained below. According to the central-limit theorem, the flux processes, $\mathcal{J}_i(t)$, are Gaussian because they are the space integrals of current densities, whose correlations are short-ranged. Stationarity implies that their Fourier transforms, Eq. \eqref{eq:GKM-3}, are normal deviates that for large $\tau$ are uncorrelated for $\omega\ne\omega'$. It follows that the sample spectrum of Eq. \eqref{eq:GKM-2}, aka the \emph{cross-periodogram}, is a collection of complex Wishart random matrices \cite{Nagar2011} that are uncorrelated among themselves for different frequencies. Now, the Schur complement of a block of dimension $P-1$ in a Wishart matrix of order $P$ is proportional to a $\chi^2$ stochastic variable \cite{Nagar2011,Bertossa2019}. We conclude that the Schur complement of the mass block, $\mathcal{S}'_E$, in the cross-periodogram given by Eq. \eqref{eq:GKM-2}, is the product of a smooth function of frequency, whose $\omega\to 0$ limit is the thermal conductivity we are after, times a set of independent, identically distributed, $\chi^2$ stochastic variables. By applying a low-pass filter to the logarithm of this quantity, one obtains a consistent estimator of the logarithm of the conductivity, as explained in Ref. \onlinecite{Bertossa2019}, a procedure that is known as \emph{cepstral analysis} in sound engineering and speech recognition applications \cite{Childers1977}.

\begin{figure}
    \centering
    \includegraphics[width=0.7\columnwidth]{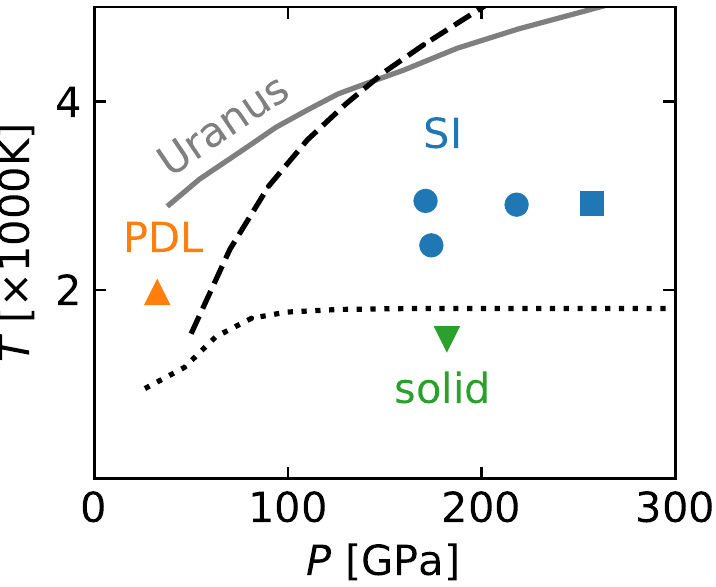}
    \caption{High-pT phase diagram of water: shown are the ice-SI (dotted) and SI-PDL curves (dashed), and Uranus' isentrope (gray solid). The symbols indicate the pT conditions at which the simulations were run.}
    \label{fig:PhaseDiagram}
\end{figure}

\begin{table*}[]
    \centering
    \begin{tabular}{
    c
    c
    c
    S[table-format=4.0(3)]
    S[table-format=3.0(2)]
    S[table-format=1.2]
    c
    S[table-format=3.1(4)]
    S[table-format=4.0(4)]
    S[table-format=4.0(4)]
    S[table-format=1.2(3)]
    c
}
        \cmidrule{1-2} \cmidrule{4-6} \cmidrule{8-12}
        \multicolumn{2}{c}{\multirow{2}*{phase}}
        & $\;$
        & $T$ & $P$  & $\rho$
        & $\;$
        & $\kappa$ & $\sigma$ & $\sigma_{NE}$ & \multicolumn{2}{c}{$\textit{D}$}
         \\
        & &
         & \si{[\kelvin]} & \si{[\giga\pascal]} & \si{[\gram\per\centi\meter\tothe{3}]}
        &
        & \si{[\watt\per(\kelvin\meter)]}
        & \si{[\siemens\per\centi\meter]}
        & \si{[\siemens\per\centi\meter]}
        & \multicolumn{2}{c}{\si{[\angstrom\tothe{2}/\pico\second]}}
         \\
        \cmidrule{1-2} \cmidrule{4-6} \cmidrule{8-12}
        ice X & \textcolor{ForestGreen}{$\blacktriangledown$} &      & 1490 \pm 50     & 182 \pm 1  & 3.52 & & 16.1 \pm 1.1  & $\mathrm{-}$ & $\mathrm{-}$ & \multicolumn{2}{c}{$\mathrm{-}$}   \\ [1.0ex]
        SI$^\mathrm{BCC}$ & \textcolor{RoyalBlue}{$\newmoon$} &      & 2470 \pm 80     & 174 \pm 2 & 3.39 & & 9.4 \pm  0.6   & 135 \pm 7 & 83 \pm 2 & 4.88 \pm 0.13 & (H)       \\
        SI$^\mathrm{BCC}$ & \textcolor{RoyalBlue}{$\newmoon$} &      & 2950 \pm 90     & 171 \pm 2 & 3.35 & & 10.7  \pm  0.7  & 180 \pm 5 & 105 \pm 2 & 7.41 \pm 0.12 & (H)        \\
        SI$^\mathrm{BCC}$ & \textcolor{RoyalBlue}{$\newmoon$}  &      & 2910 \pm 90     & 218  \pm 2 & 3.61 & & 9.9 \pm 0.7   & 198 \pm 9 & 114 \pm 2 & 7.38 \pm 0.16  & (H)     \\
        SI$^\mathrm{FCC}$ & \textcolor{RoyalBlue}{$\blacksquare$} &   & 2920 \pm 90     & 257 \pm 2 & 3.82    & & 12.8  \pm  1.0  & 256 \pm 8 & 117 \pm 2 & 7.17 \pm 0.13 & (H)       \\ [1ex]
        {PDL} & \textcolor{BurntOrange}{$\blacktriangle$}   &    & 1970 \pm 60     & 33 \pm 1 & 2.04   & & 4.1  \pm  0.3   & 42 \pm 3 & & 3.10 \pm 0.03 & (H)   \\
         & & & & & & & & & & 0.92 \pm 0.02 & (O)  \\ [0.5ex]
        \cmidrule{1-2} \cmidrule{4-6} \cmidrule{8-12}
    \end{tabular}
    \caption{{Summary of our results. $T$, $P$, and $\rho$ indicate temperature, pressure, and density, respectively. $\kappa$, $\sigma$, and $D$ are thermal and electrical conductivities and atomic diffusivities, respectively. $\sigma_{NE}$ is the value of the electric conductivity obtained from the Nernst-Einstein relation, Eq. \eqref{eq:Nernst}}.}
    \label{tab:NVE_results}
\end{table*}

\subsection{Simulations}
The heat and charge transport properties of different (solid, PDL, and SI) phases of water in the 1,000-3,000 K and 30-250 GPa pT range have been explored by Car-Parrinello (CP) \emph{ab initio} NVE molecular dynamics \cite{Car1985}, {using the \textsc{Quantum ESPRESSO} suite of computer codes \cite{Giannozzi2009,Giannozzi2017}}. We believe that the CP Lagrangian formalism is particularly fit for transport simulations because the accurate conservation of the (extended) total energy allows one to generate long and stable trajectories without using thermostats. Figure \ref{fig:PhaseDiagram} shows the phase diagram of water in such pT range. The SI-{PDL} (dashed) and ice-SI (dotted) phase boundaries are obtained from state-of-the-art shock-compression experiments \cite{Millot2018}; Uranus' isentrope (solid gray) from \emph{ab initio simulations} \cite{Redmer2011} is also reported. We have verified that a body-centered-cubic (BCC) to face-centered-cubic (FCC) transition in the oxygen lattice occurs for the SI phase at $P \approx 240$  GPa and $T\approx$ 3000 K, in accordance with {recent theoretical} \cite{Sun2015} {and experimental findings \cite{Millot2019}}. We then ran three simulations for the BCC-SI phase (blue circles) and one for the FCC-SI one (blue square). We also ran a simulation for solid ice X (green triangle) and a simulation for the PDL (orange triangle) at pT conditions where the fraction of dissociated molecules is $\sim$10\% \cite{French2010}. We have explicitly checked that the electron energy gap computed along the various MD trajectories is always way larger than $k_BT$, thus ruling out any direct electronic contributions to heat and charge transport. All the technical details of the simulations are reported in the Supplementary Information \cite{SupplMat}. {Our results are summarized in Table \ref{tab:NVE_results}.}

\section{Discussion}

\begin{figure}[t]
    \centering
    \includegraphics[width=0.85\columnwidth]{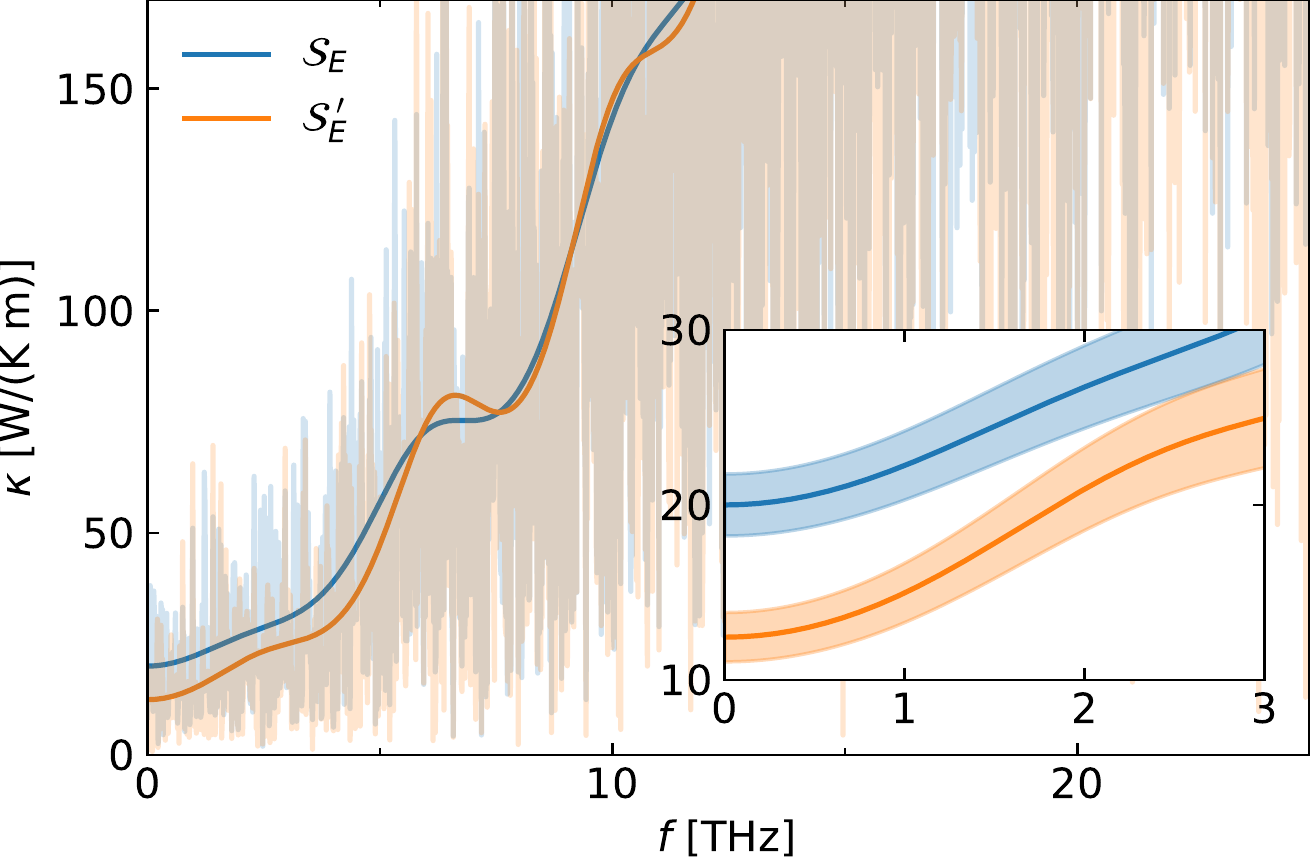}
    \caption{Energy-flux power spectra for FCC-SI water at average $T = 2920\un{K}$ and $P = 257\un{GPa}$. Blue: plain periodogram, $\mathcal{S}_E$. Orange: reduced periodogram, $\mathcal{S}'_E$, computed at vanishing mass fluxes, Eq. \eqref{eq:multi_kappa}. The thick lines are the filtered spectra obtained via cepstral analysis. Inset: low-frequency zoom of with their estimated uncertainties.}
    \label{fig:FCC}
\end{figure}

We start the discussion of our results by highlighting the importance of a multi-component analysis of the heat- and mass-flux time series resulting from our simulations. In Fig. \ref{fig:FCC} we display the power spectrum of the energy flux of FCC-SI water at an average temperature $T = 2920 \pm 90 \un{K}$ and pressure $P = 257 \pm 2 \un{GPa}$, evaluated according to two different prescriptions: blue lines refer to the plain spectrum of the energy flux computed within density-functional theory using the formulation of  Ref. \onlinecite{Marcolongo2016}; orange lines indicate the ``residual spectrum'' computed by assuming that the mass flux vanishes, according to Eq. \eqref{eq:multi_kappa}. The sample power spectra (the ``periodograms'') are displayed with faint lines, whereas those subject to cepstral filtering are displayed with thick lines; the latter are zoomed-in at low frequency and displayed in the inset, together with their statistical uncertainties. By looking at the zero-frequency value of the spectrum, cepstral analysis gives $\kappa = 20 \pm 2 \un{W/(Km)}$, and $\kappa = 13 \pm 2 \un{W/(Km)}$ neglecting and accounting for the interaction with the H mass flux, respectively. {In effectively one-component systems, statistical analysis can be greatly facilitated by fixing a suitably defined optimal gauge for the diffusing current \cite{antani2020}.} Since SI is a truly bi-component system, a bi-variate analysis is indeed needed to account for the interaction between different conserved fluxes and for a correct estimate of $\kappa$: considering the time series of the energy flux alone---as if the system were one-component---would overestimate the heat conductivity by 80\%.

Convective invariance can also be leveraged to reduce the statistical noise, and thus the uncertainty, on the estimated value of $\kappa$, as explained in Ref. \onlinecite{Bertossa2019}. The addition of one or more components to the set of conserved fluxes to be analysed decreases the total power of the reduced spectrum without affecting its value at zero frequency, thus making it smoother and the low-pass cepstral filter more efficacious. By adopting the adiabatic electron current as an additional flux, one obtains the refined result: $\kappa = 12.8 \pm 1.0 \un{W/(K m)}$. Further details on the statistical analysis of our data can be found in the Supplementary Information \cite{SupplMat}.

\begin{figure}
    \centering
    \includegraphics[width=0.85\columnwidth]{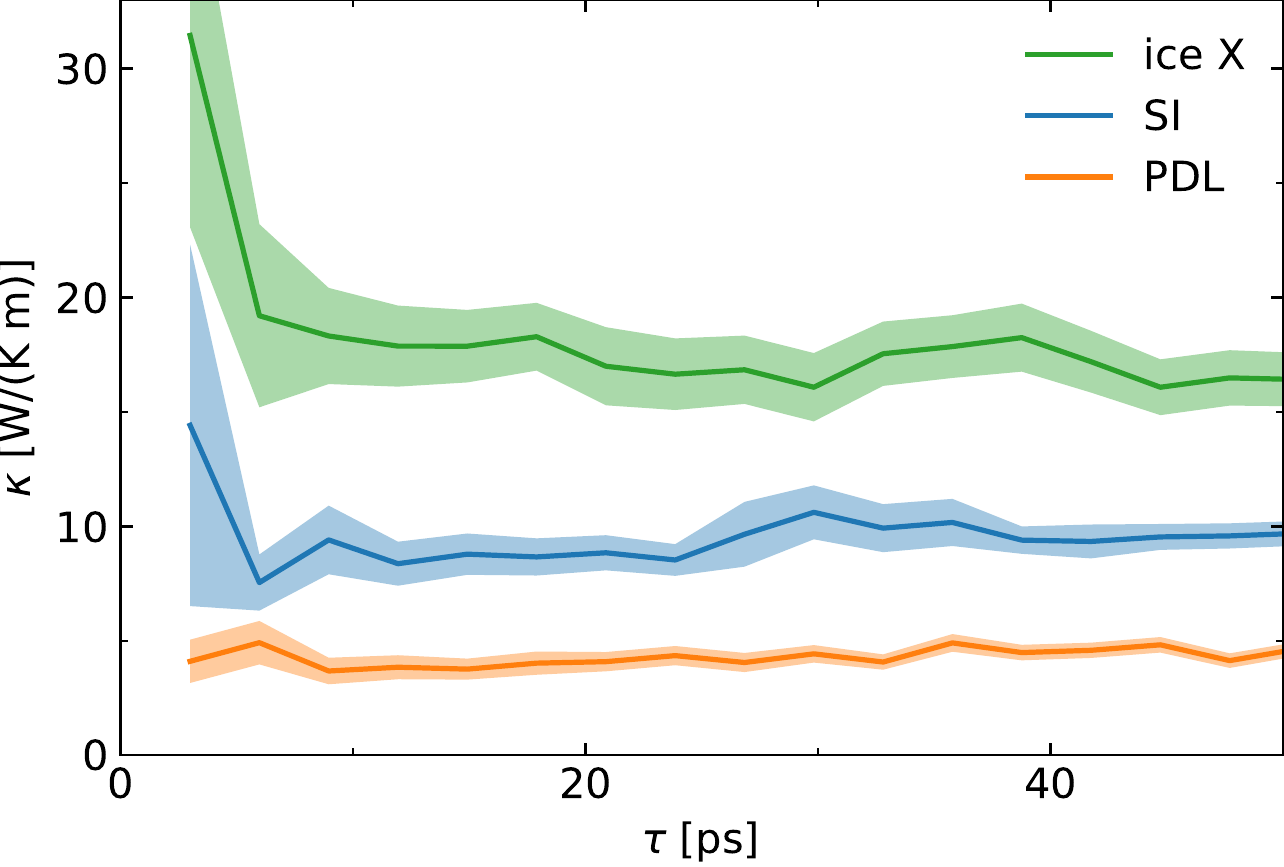}
    \caption{Thermal conductivity as a function of the length of the AIMD trajectory: solid ice X (green), BCC-SI ($T=2470$ \si{K}, $P=174$ GPa,  blue), and PDL (orange) phases of water, see Table \ref{tab:NVE_results}. The shaded areas indicate the estimated statistical uncertainty.}
    \label{fig:kVSt}
\end{figure}

Multi-component cepstral analysis, which has been performed using the \texttt{thermocepstrum} code \cite{thermocepstrum}, allows us to obtain  accurate transport coefficients from relatively short AIMD trajectories, particularly for the strongly anharmonic exotic phases of water occurring at the high pT conditions of interest here. Figure \ref{fig:kVSt} shows the values and the statistical uncertainties of the heat conductivity of different phases of water as a function of the length of the (reduced) energy-flux time series from which they are estimated.
These data show that well-converged results with an uncertainty of $\approx 15\%$ are obtained with trajectories as short as 10-20 ps. Not surprisingly, the more crystalline a phase is, the larger the uncertainty for a same trajectory length  ($\text{ice X}>\text{{SI}}>\text{{PDL}}$), due to the larger residual harmonicity of the structure. {We stress that cepstral analysis is a \emph{self-averaging technique}, in that the statistical error affecting the estimated {conductivities} can be accurately estimated and systematically reduced by increasing the length of the simulation, thus avoiding the need to average over different MD trajectories. Nonetheless, isotropy allows one to consider the three Cartesian components of the fluxes as different samples of a same process: the {spectra have} been thus averaged over {Cartesian components}.}

Our results are summarized in Table \ref{tab:NVE_results}. In the pT-conditions examined here, the thermal conductivity of solid ice X is larger than that of the SI phase, which is itself larger than in PDL water. This is expected, again due to the decreasing level of harmonicity in going from a crystalline to a partially liquid and eventually fully liquid phase. We did not observe a significant dependence of $\kappa$ upon the temperature for the SI phase in the explored range.
The FCC-SI water has slightly larger heat conductivity than BCC-SI.

Pioneering AIMD simulations of charge transport in PDL water \cite{French2011} revealed that, rather unexpectedly, a classical model of charge conduction where hydrogen and oxygen ions carry an integer charge whose magnitudes equal their formal oxidation numbers ($q_\mathrm{H} = +1$ and $q_\mathrm{O} = -2$) yields the same conductivity that would be obtained from the exact quantum-mechanical expression of the electric current, based on Born's effective charges. This surprising finding was {given a solid theoretical foundation} in a recent paper of ours where it was shown to result from the combined effects of gauge invariance of transport coefficients and topological quantization of adiabatic charge transport \cite{Grasselli2019}. Leveraging this result, we computed the electrical conductivity from the cepstral analysis of the classical charge flux, defined as:
\begin{equation}
    \mathcal{J}_{Z} = \frac{1}{\Omega} \left( q_\mathrm{H} \sum_{n\in \mathrm{H}}  \mathcal{V}_{n} +  q_\mathrm{O} \sum_{n \in \mathrm{O}} \mathcal{V}_{n} \right), \label{eq:classical_current}
\end{equation}
where the $\mathcal V$'s are ionic velocities.

The electrical conductivities resulting from our simulations are reported in Table \ref{tab:NVE_results}. The {data} tagged with the ``$NE$'' subscript are obtained using the Nernst-Einstein equation \cite{Marcolongo2017}, which neglects all inter-ionic correlations and that in the one-component case reads:
\begin{equation}
    \sigma_{NE}= \frac{e^2 q^2_\mathrm{H} N_\mathrm{H} D_\mathrm{H}}{\Omega k_B T},
    \label{eq:Nernst}
\end{equation}
where $N_H$ and $D_H$ are the number of hydrogen atoms and their diffusivity, respectively. In the case of PDL, Eq. \eqref{eq:Nernst} hardly applies, as it would depend on too large a number of parameters (the concentrations, life-times, and diffusivities of the various ionic charge carriers). Our results are consistent with previous theoretical estimates \cite{French2011,Sun2015}, as well as with the experimental data obtained from electrical impedance measurements along the liquid or pre-compressed Hugoniot \cite{Mitchell1982,Chau2001,Yakushev2000}, summarized in Fig. 4 of Ref. \onlinecite{Millot2018}: $\sigma \sim 150 \un{S/cm}$ for the SI phase in the range 100-150 GPa and 2000-3000 K; and $\sigma \sim 30 \un{S/cm}$ for the PDL phase at $\approx 30 \un{GPa}$ and $2000 \un{K}$. Two important trends emerge from our results. First, the NE relation severely underestimates the conductivity in {SI} water, as already observed in other {SI} systems \cite{Marcolongo2017}. At variance with these findings, when charge carriers of opposite signs coexist in an electrolyte, the short-range correlations among them may screen the amount of transported charge, thus determining a \emph{decrease} of the electric conductivity with respect to the predictions of the NE approximation \cite{France-Lanord2019}.
{In the second place, the electrical conductivity {in} the FCC {SI} phase is sensitively larger than {in} the BCC one, in contrast to the opposite trend displayed by hydrogen \textit{diffusivity}, which  are instead slightly smaller in the FCC phase, thus resulting in comparable predictions for the two phases of the NE approximation ($\sigma_{NE}$). The lesser ability of the NE approximation to predict the conductivity in the FCC than in the BCC phase indicates a stronger effect of inter-ionic correlations in the former case: the higher energy barriers for a single proton hop in FCC---due to its larger packing density \cite{Wilson2013}, and resulting in a slightly smaller ionic diffusivity---may be effectively decreased by a cooperative motion of \textit{two or more} protons (as already observed for the carrier dynamics in solid-state electrolytes \cite{He2017}), and thus lead to an overall larger electrical conductivity.}

\section{Conclusions}
In this paper we have reported on the {first theoretically rigorous and numerically accurate evaluation of the thermal and electric conductivities of various phases of water occurring at the pressure and temperature conditions to be found} in the interior of ice giant planets, made possible by recent advances in transport theory and data analysis. {In the case of the heat conductivity, our results set a reference in the wide range of values used in evolution models of Uranus and Neptune \cite{Podolak2019} or given by recent MD-based estimates on dissociating water \cite{French2019}, and their moderate values point towards more efficient trapping of heat in the deep interior of these planets. These results have been instrumental in the development of a novel model of the thermal evolution of Uranus, featuring a frozen core and an anomalously low heat flow, resulting in the observed low luminosity of this planet \cite{noi2020}. Finally, the electrical conductivity that we find for {SI} ice is far larger than assumed in previous models of the generation of the magnetic fields in Uranus and Neptune \cite{Stanley2006}. Since {SI} ice is likely to dominate the deeper sluggish layer that underlies the shallow fluid outer layer in which the magnetic field is produced, the large electrical conductivity of the {SI} phase can have a substantial impact on the geometry and time evolution of the magnetic field of these planets.}

\section{Data availability}
The data that support the plots and relevant results within this paper are available on the Materials Cloud Platform at \url{https://doi.org/10.24435/materialscloud:hn-6f}.

\section{Acknowledgments}

\begin{acknowledgments} This work was partially funded by the EU through the \textsc{MaX} Centre of Excellence for supercomputing applications (Project No. 824143), {and by the US National Science Foundation under grant EAR-1853388}. We are grateful to Riccardo Bertossa {and Davide Tisi} for valuable {discussions and assistance}. \end{acknowledgments}

\section{Author contributions}
F.G., L.S., and S.B. contributed to conceive this research, perform the simulations, analyze the results, and write the paper.

\section{Competing interests}
The authors declare no competing interests.

\section{Code availability statement}
Computer codes are available and freely downloadable from the \textsc{Quantum ESPRESSO} site and the \texttt{Thermocepstrum} GitHub page referenced below.



\begin{thebibliography}{79}%
\makeatletter
\providecommand \@ifxundefined [1]{%
 \@ifx{#1\undefined}
}%
\providecommand \@ifnum [1]{%
 \ifnum #1\expandafter \@firstoftwo
 \else \expandafter \@secondoftwo
 \fi
}%
\providecommand \@ifx [1]{%
 \ifx #1\expandafter \@firstoftwo
 \else \expandafter \@secondoftwo
 \fi
}%
\providecommand \natexlab [1]{#1}%
\providecommand \enquote  [1]{``#1''}%
\providecommand \bibnamefont  [1]{#1}%
\providecommand \bibfnamefont [1]{#1}%
\providecommand \citenamefont [1]{#1}%
\providecommand \href@noop [0]{\@secondoftwo}%
\providecommand \href [0]{\begingroup \@sanitize@url \@href}%
\providecommand \@href[1]{\@@startlink{#1}\@@href}%
\providecommand \@@href[1]{\endgroup#1\@@endlink}%
\providecommand \@sanitize@url [0]{\catcode `\\12\catcode `\$12\catcode
  `\&12\catcode `\#12\catcode `\^12\catcode `\_12\catcode `\%12\relax}%
\providecommand \@@startlink[1]{}%
\providecommand \@@endlink[0]{}%
\providecommand \url  [0]{\begingroup\@sanitize@url \@url }%
\providecommand \@url [1]{\endgroup\@href {#1}{\urlprefix }}%
\providecommand \urlprefix  [0]{URL }%
\providecommand \Eprint [0]{\href }%
\providecommand \doibase [0]{http://dx.doi.org/}%
\providecommand \selectlanguage [0]{\@gobble}%
\providecommand \bibinfo  [0]{\@secondoftwo}%
\providecommand \bibfield  [0]{\@secondoftwo}%
\providecommand \translation [1]{[#1]}%
\providecommand \BibitemOpen [0]{}%
\providecommand \bibitemStop [0]{}%
\providecommand \bibitemNoStop [0]{.\EOS\space}%
\providecommand \EOS [0]{\spacefactor3000\relax}%
\providecommand \BibitemShut  [1]{\csname bibitem#1\endcsname}%
\let\auto@bib@innerbib\@empty
\bibitem [{\citenamefont {Lodders}(2003)}]{Lodders2003}%
  \BibitemOpen
  \bibfield  {author} {\bibinfo {author} {\bibfnamefont {Katharina}\
  \bibnamefont {Lodders}},\ }\bibfield  {title} {\enquote {\bibinfo {title}
  {Solar system abundances and condensation temperatures of the elements},}\
  }\href@noop {} {\bibfield  {journal} {\bibinfo  {journal} {The Astrophysical
  Journal}\ }\textbf {\bibinfo {volume} {591}},\ \bibinfo {pages} {1220}
  (\bibinfo {year} {2003})}\BibitemShut {NoStop}%
\bibitem [{\citenamefont {Nimmo}\ \emph {et~al.}(2007)\citenamefont {Nimmo},
  \citenamefont {Spencer}, \citenamefont {Pappalardo},\ and\ \citenamefont
  {Mullen}}]{Nimmo2007}%
  \BibitemOpen
  \bibfield  {author} {\bibinfo {author} {\bibfnamefont {Francis}\ \bibnamefont
  {Nimmo}}, \bibinfo {author} {\bibfnamefont {JR}~\bibnamefont {Spencer}},
  \bibinfo {author} {\bibfnamefont {RT}~\bibnamefont {Pappalardo}}, \ and\
  \bibinfo {author} {\bibfnamefont {ME}~\bibnamefont {Mullen}},\ }\bibfield
  {title} {\enquote {\bibinfo {title} {Shear heating as the origin of the
  plumes and heat flux on enceladus},}\ }\href@noop {} {\bibfield  {journal}
  {\bibinfo  {journal} {Nature}\ }\textbf {\bibinfo {volume} {447}},\ \bibinfo
  {pages} {289--291} (\bibinfo {year} {2007})}\BibitemShut {NoStop}%
\bibitem [{\citenamefont {Kivelson}\ \emph {et~al.}(1996)\citenamefont
  {Kivelson}, \citenamefont {Khurana}, \citenamefont {Russell}, \citenamefont
  {Walker}, \citenamefont {Warnecke}, \citenamefont {Coroniti}, \citenamefont
  {Polanskey}, \citenamefont {Southwood},\ and\ \citenamefont
  {Schubert}}]{Kivelson1996}%
  \BibitemOpen
  \bibfield  {author} {\bibinfo {author} {\bibfnamefont {MG}~\bibnamefont
  {Kivelson}}, \bibinfo {author} {\bibfnamefont {KK}~\bibnamefont {Khurana}},
  \bibinfo {author} {\bibfnamefont {CT}~\bibnamefont {Russell}}, \bibinfo
  {author} {\bibfnamefont {RJ}~\bibnamefont {Walker}}, \bibinfo {author}
  {\bibfnamefont {J}~\bibnamefont {Warnecke}}, \bibinfo {author} {\bibfnamefont
  {FV}~\bibnamefont {Coroniti}}, \bibinfo {author} {\bibfnamefont
  {C}~\bibnamefont {Polanskey}}, \bibinfo {author} {\bibfnamefont
  {DJ}~\bibnamefont {Southwood}}, \ and\ \bibinfo {author} {\bibfnamefont
  {G}~\bibnamefont {Schubert}},\ }\bibfield  {title} {\enquote {\bibinfo
  {title} {Discovery of ganymede's magnetic field by the galileo spacecraft},}\
  }\href@noop {} {\bibfield  {journal} {\bibinfo  {journal} {Nature}\ }\textbf
  {\bibinfo {volume} {384}},\ \bibinfo {pages} {537--541} (\bibinfo {year}
  {1996})}\BibitemShut {NoStop}%
\bibitem [{\citenamefont {Nettelmann}\ \emph {et~al.}(2013)\citenamefont
  {Nettelmann}, \citenamefont {Helled}, \citenamefont {Fortney},\ and\
  \citenamefont {Redmer}}]{Nettelmann2013}%
  \BibitemOpen
  \bibfield  {author} {\bibinfo {author} {\bibfnamefont {N}~\bibnamefont
  {Nettelmann}}, \bibinfo {author} {\bibfnamefont {R}~\bibnamefont {Helled}},
  \bibinfo {author} {\bibfnamefont {JJ}~\bibnamefont {Fortney}}, \ and\
  \bibinfo {author} {\bibfnamefont {R}~\bibnamefont {Redmer}},\ }\bibfield
  {title} {\enquote {\bibinfo {title} {New indication for a dichotomy in the
  interior structure of uranus and neptune from the application of modified
  shape and rotation data},}\ }\href@noop {} {\bibfield  {journal} {\bibinfo
  {journal} {Planetary and Space Science}\ }\textbf {\bibinfo {volume} {77}},\
  \bibinfo {pages} {143--151} (\bibinfo {year} {2013})}\BibitemShut {NoStop}%
\bibitem [{\citenamefont {Cavazzoni}\ \emph {et~al.}(1999)\citenamefont
  {Cavazzoni}, \citenamefont {Chiarotti}, \citenamefont {Scandolo},
  \citenamefont {Tosatti}, \citenamefont {Bernasconi},\ and\ \citenamefont
  {Parrinello}}]{Cavazzoni1999}%
  \BibitemOpen
  \bibfield  {author} {\bibinfo {author} {\bibfnamefont {C.}~\bibnamefont
  {Cavazzoni}}, \bibinfo {author} {\bibfnamefont {G.~L.}\ \bibnamefont
  {Chiarotti}}, \bibinfo {author} {\bibfnamefont {S.}~\bibnamefont {Scandolo}},
  \bibinfo {author} {\bibfnamefont {E.}~\bibnamefont {Tosatti}}, \bibinfo
  {author} {\bibfnamefont {M.}~\bibnamefont {Bernasconi}}, \ and\ \bibinfo
  {author} {\bibfnamefont {M.}~\bibnamefont {Parrinello}},\ }\bibfield  {title}
  {\enquote {\bibinfo {title} {{Superionic and Metallic States of Water and
  Ammonia at Giant Planet Conditions.}}}\ }\href {\doibase
  10.1126/science.283.5398.44} {\bibfield  {journal} {\bibinfo  {journal}
  {Science}\ }\textbf {\bibinfo {volume} {283}},\ \bibinfo {pages} {44--46}
  (\bibinfo {year} {1999})}\BibitemShut {NoStop}%
\bibitem [{\citenamefont {Millot}\ \emph {et~al.}(2018)\citenamefont {Millot},
  \citenamefont {Hamel}, \citenamefont {Rygg}, \citenamefont {Celliers},
  \citenamefont {Collins}, \citenamefont {Coppari}, \citenamefont
  {Fratanduono}, \citenamefont {Jeanloz}, \citenamefont {Swift},\ and\
  \citenamefont {Eggert}}]{Millot2018}%
  \BibitemOpen
  \bibfield  {author} {\bibinfo {author} {\bibfnamefont {Marius}\ \bibnamefont
  {Millot}}, \bibinfo {author} {\bibfnamefont {Sebastien}\ \bibnamefont
  {Hamel}}, \bibinfo {author} {\bibfnamefont {J~Ryan}\ \bibnamefont {Rygg}},
  \bibinfo {author} {\bibfnamefont {Peter~M}\ \bibnamefont {Celliers}},
  \bibinfo {author} {\bibfnamefont {Gilbert~W}\ \bibnamefont {Collins}},
  \bibinfo {author} {\bibfnamefont {Federica}\ \bibnamefont {Coppari}},
  \bibinfo {author} {\bibfnamefont {Dayne~E}\ \bibnamefont {Fratanduono}},
  \bibinfo {author} {\bibfnamefont {Raymond}\ \bibnamefont {Jeanloz}}, \bibinfo
  {author} {\bibfnamefont {Damian~C}\ \bibnamefont {Swift}}, \ and\ \bibinfo
  {author} {\bibfnamefont {Jon~H}\ \bibnamefont {Eggert}},\ }\bibfield  {title}
  {\enquote {\bibinfo {title} {Experimental evidence for superionic water ice
  using shock compression},}\ }\href@noop {} {\bibfield  {journal} {\bibinfo
  {journal} {Nature Physics}\ }\textbf {\bibinfo {volume} {14}},\ \bibinfo
  {pages} {297--302} (\bibinfo {year} {2018})}\BibitemShut {NoStop}%
\bibitem [{\citenamefont {Stanley}\ and\ \citenamefont
  {Bloxham}(2004)}]{Stanley2004}%
  \BibitemOpen
  \bibfield  {author} {\bibinfo {author} {\bibfnamefont {Sabine}\ \bibnamefont
  {Stanley}}\ and\ \bibinfo {author} {\bibfnamefont {Jeremy}\ \bibnamefont
  {Bloxham}},\ }\bibfield  {title} {\enquote {\bibinfo {title}
  {Convective-region geometry as the cause of uranus' and neptune's unusual
  magnetic fields},}\ }\href@noop {} {\bibfield  {journal} {\bibinfo  {journal}
  {Nature}\ }\textbf {\bibinfo {volume} {428}},\ \bibinfo {pages} {151--153}
  (\bibinfo {year} {2004})}\BibitemShut {NoStop}%
\bibitem [{\citenamefont {Zeng}\ \emph {et~al.}(2019)\citenamefont {Zeng},
  \citenamefont {Jacobsen}, \citenamefont {Sasselov}, \citenamefont {Petaev},
  \citenamefont {Vanderburg}, \citenamefont {Lopez-Morales}, \citenamefont
  {Perez-Mercader}, \citenamefont {Mattsson}, \citenamefont {Li}, \citenamefont
  {Heising} \emph {et~al.}}]{Zeng2019}%
  \BibitemOpen
  \bibfield  {author} {\bibinfo {author} {\bibfnamefont {Li}~\bibnamefont
  {Zeng}}, \bibinfo {author} {\bibfnamefont {Stein~B}\ \bibnamefont
  {Jacobsen}}, \bibinfo {author} {\bibfnamefont {Dimitar~D}\ \bibnamefont
  {Sasselov}}, \bibinfo {author} {\bibfnamefont {Michail~I}\ \bibnamefont
  {Petaev}}, \bibinfo {author} {\bibfnamefont {Andrew}\ \bibnamefont
  {Vanderburg}}, \bibinfo {author} {\bibfnamefont {Mercedes}\ \bibnamefont
  {Lopez-Morales}}, \bibinfo {author} {\bibfnamefont {Juan}\ \bibnamefont
  {Perez-Mercader}}, \bibinfo {author} {\bibfnamefont {Thomas~R}\ \bibnamefont
  {Mattsson}}, \bibinfo {author} {\bibfnamefont {Gongjie}\ \bibnamefont {Li}},
  \bibinfo {author} {\bibfnamefont {Matthew~Z}\ \bibnamefont {Heising}},  \emph
  {et~al.},\ }\bibfield  {title} {\enquote {\bibinfo {title} {Growth model
  interpretation of planet size distribution},}\ }\href@noop {} {\bibfield
  {journal} {\bibinfo  {journal} {Proceedings of the National Academy of
  Sciences}\ }\textbf {\bibinfo {volume} {116}},\ \bibinfo {pages} {9723--9728}
  (\bibinfo {year} {2019})}\BibitemShut {NoStop}%
\bibitem [{\citenamefont {Fortney}\ and\ \citenamefont
  {Nettelmann}(2010)}]{Fortney2010}%
  \BibitemOpen
  \bibfield  {author} {\bibinfo {author} {\bibfnamefont {Jonathan~J.}\
  \bibnamefont {Fortney}}\ and\ \bibinfo {author} {\bibfnamefont {Nadine}\
  \bibnamefont {Nettelmann}},\ }\bibfield  {title} {\enquote {\bibinfo {title}
  {The interior structure, composition, and evolution of~giant planets},}\
  }\href {\doibase 10.1007/s11214-009-9582-x} {\bibfield  {journal} {\bibinfo
  {journal} {Space Sci. Rev.}\ }\textbf {\bibinfo {volume} {152}},\ \bibinfo
  {pages} {423--447} (\bibinfo {year} {2010})}\BibitemShut {NoStop}%
\bibitem [{\citenamefont {Hubbard}\ \emph {et~al.}(1995)\citenamefont
  {Hubbard}, \citenamefont {Podolak},\ and\ \citenamefont
  {Stevenson}}]{Hubbard:1995}%
  \BibitemOpen
  \bibfield  {author} {\bibinfo {author} {\bibfnamefont {W.B}\ \bibnamefont
  {Hubbard}}, \bibinfo {author} {\bibfnamefont {M.}~\bibnamefont {Podolak}}, \
  and\ \bibinfo {author} {\bibfnamefont {D.J.}\ \bibnamefont {Stevenson}},\
  }\bibfield  {title} {\enquote {\bibinfo {title} {The interior of neptune},}\
  }in\ \href@noop {} {\emph {\bibinfo {booktitle} {Neptune and Triton}}},\
  \bibinfo {editor} {edited by\ \bibinfo {editor} {\bibfnamefont {Dale~P.}\
  \bibnamefont {Cruikshank}}}\ (\bibinfo  {publisher} {University of Arizona
  Press},\ \bibinfo {year} {1995})\ p.\ \bibinfo {pages}
  {109–138}\BibitemShut {NoStop}%
\bibitem [{\citenamefont {Helled}\ \emph {et~al.}(2020)\citenamefont {Helled},
  \citenamefont {Nettelmann},\ and\ \citenamefont {Guillot}}]{Helled2020}%
  \BibitemOpen
  \bibfield  {author} {\bibinfo {author} {\bibfnamefont {Ravit}\ \bibnamefont
  {Helled}}, \bibinfo {author} {\bibfnamefont {Nadine}\ \bibnamefont
  {Nettelmann}}, \ and\ \bibinfo {author} {\bibfnamefont {Tristan}\
  \bibnamefont {Guillot}},\ }\bibfield  {title} {\enquote {\bibinfo {title}
  {{Uranus and Neptune: Origin, Evolution and Internal Structure}},}\ }\href
  {\doibase 10.1007/s11214-020-00660-3} {\bibfield  {journal} {\bibinfo
  {journal} {Space Science Reviews}\ }\textbf {\bibinfo {volume} {216}}
  (\bibinfo {year} {2020}),\ 10.1007/s11214-020-00660-3},\ \Eprint
  {http://arxiv.org/abs/1909.04891} {arXiv:1909.04891} \BibitemShut {NoStop}%
\bibitem [{\citenamefont {Mitchell}\ and\ \citenamefont
  {Nellis}(1982)}]{Mitchell1982}%
  \BibitemOpen
  \bibfield  {author} {\bibinfo {author} {\bibfnamefont {AC}~\bibnamefont
  {Mitchell}}\ and\ \bibinfo {author} {\bibfnamefont {WJ}~\bibnamefont
  {Nellis}},\ }\bibfield  {title} {\enquote {\bibinfo {title} {Equation of
  state and electrical conductivity of water and ammonia shocked to the 100 gpa
  (1 mbar) pressure range},}\ }\href@noop {} {\bibfield  {journal} {\bibinfo
  {journal} {The Journal of Chemical Physics}\ }\textbf {\bibinfo {volume}
  {76}},\ \bibinfo {pages} {6273--6281} (\bibinfo {year} {1982})}\BibitemShut
  {NoStop}%
\bibitem [{\citenamefont {Yakushev}\ \emph {et~al.}(2000)\citenamefont
  {Yakushev}, \citenamefont {Postnov}, \citenamefont {Fortov},\ and\
  \citenamefont {Yakysheva}}]{Yakushev2000}%
  \BibitemOpen
  \bibfield  {author} {\bibinfo {author} {\bibfnamefont {VV}~\bibnamefont
  {Yakushev}}, \bibinfo {author} {\bibfnamefont {VI}~\bibnamefont {Postnov}},
  \bibinfo {author} {\bibfnamefont {VE}~\bibnamefont {Fortov}}, \ and\ \bibinfo
  {author} {\bibfnamefont {TI}~\bibnamefont {Yakysheva}},\ }\bibfield  {title}
  {\enquote {\bibinfo {title} {Electrical conductivity of water during
  quasi-isentropic compression to 130 gpa},}\ }\href@noop {} {\bibfield
  {journal} {\bibinfo  {journal} {Journal of Experimental and Theoretical
  Physics}\ }\textbf {\bibinfo {volume} {90}},\ \bibinfo {pages} {617--622}
  (\bibinfo {year} {2000})}\BibitemShut {NoStop}%
\bibitem [{\citenamefont {Chau}\ \emph {et~al.}(2001)\citenamefont {Chau},
  \citenamefont {Mitchell}, \citenamefont {Minich},\ and\ \citenamefont
  {Nellis}}]{Chau2001}%
  \BibitemOpen
  \bibfield  {author} {\bibinfo {author} {\bibfnamefont {R}~\bibnamefont
  {Chau}}, \bibinfo {author} {\bibfnamefont {AC}~\bibnamefont {Mitchell}},
  \bibinfo {author} {\bibfnamefont {RW}~\bibnamefont {Minich}}, \ and\ \bibinfo
  {author} {\bibfnamefont {WJ}~\bibnamefont {Nellis}},\ }\bibfield  {title}
  {\enquote {\bibinfo {title} {Electrical conductivity of water compressed
  dynamically to pressures of 70--180 gpa (0.7--1.8 mbar)},}\ }\href@noop {}
  {\bibfield  {journal} {\bibinfo  {journal} {The Journal of Chemical Physics}\
  }\textbf {\bibinfo {volume} {114}},\ \bibinfo {pages} {1361--1365} (\bibinfo
  {year} {2001})}\BibitemShut {NoStop}%
\bibitem [{\citenamefont {Rozsa}\ \emph {et~al.}(2018)\citenamefont {Rozsa},
  \citenamefont {Pan}, \citenamefont {Giberti},\ and\ \citenamefont
  {Galli}}]{Rozsa2018}%
  \BibitemOpen
  \bibfield  {author} {\bibinfo {author} {\bibfnamefont {Viktor}\ \bibnamefont
  {Rozsa}}, \bibinfo {author} {\bibfnamefont {Ding}\ \bibnamefont {Pan}},
  \bibinfo {author} {\bibfnamefont {Federico}\ \bibnamefont {Giberti}}, \ and\
  \bibinfo {author} {\bibfnamefont {Giulia}\ \bibnamefont {Galli}},\ }\bibfield
   {title} {\enquote {\bibinfo {title} {Ab initio spectroscopy and ionic
  conductivity of water under earth mantle conditions},}\ }\href@noop {}
  {\bibfield  {journal} {\bibinfo  {journal} {Proceedings of the National
  Academy of Sciences}\ }\textbf {\bibinfo {volume} {115}},\ \bibinfo {pages}
  {6952--6957} (\bibinfo {year} {2018})}\BibitemShut {NoStop}%
\bibitem [{\citenamefont {Sun}\ \emph {et~al.}(2015)\citenamefont {Sun},
  \citenamefont {Clark}, \citenamefont {Torquato},\ and\ \citenamefont
  {Car}}]{Sun2015}%
  \BibitemOpen
  \bibfield  {author} {\bibinfo {author} {\bibfnamefont {Jiming}\ \bibnamefont
  {Sun}}, \bibinfo {author} {\bibfnamefont {Bryan~K}\ \bibnamefont {Clark}},
  \bibinfo {author} {\bibfnamefont {Salvatore}\ \bibnamefont {Torquato}}, \
  and\ \bibinfo {author} {\bibfnamefont {Roberto}\ \bibnamefont {Car}},\
  }\bibfield  {title} {\enquote {\bibinfo {title} {The phase diagram of
  high-pressure superionic ice},}\ }\href@noop {} {\bibfield  {journal}
  {\bibinfo  {journal} {Nature communications}\ }\textbf {\bibinfo {volume}
  {6}},\ \bibinfo {pages} {8156} (\bibinfo {year} {2015})}\BibitemShut
  {NoStop}%
\bibitem [{\citenamefont {French}\ \emph {et~al.}(2011)\citenamefont {French},
  \citenamefont {Hamel},\ and\ \citenamefont {Redmer}}]{French2011}%
  \BibitemOpen
  \bibfield  {author} {\bibinfo {author} {\bibfnamefont {Martin}\ \bibnamefont
  {French}}, \bibinfo {author} {\bibfnamefont {Sebastien}\ \bibnamefont
  {Hamel}}, \ and\ \bibinfo {author} {\bibfnamefont {Ronald}\ \bibnamefont
  {Redmer}},\ }\bibfield  {title} {\enquote {\bibinfo {title} {Dynamical
  screening and ionic conductivity in water from ab initio simulations},}\
  }\href {\doibase 10.1103/PhysRevLett.107.185901} {\bibfield  {journal}
  {\bibinfo  {journal} {Phys. Rev. Lett.}\ }\textbf {\bibinfo {volume} {107}},\
  \bibinfo {pages} {185901} (\bibinfo {year} {2011})}\BibitemShut {NoStop}%
\bibitem [{\citenamefont {Green}(1952)}]{Green1952}%
  \BibitemOpen
  \bibfield  {author} {\bibinfo {author} {\bibfnamefont {Melville~S.}\
  \bibnamefont {Green}},\ }\bibfield  {title} {\enquote {\bibinfo {title}
  {Markoff random processes and the statistical mechanics of time‐dependent
  phenomena.}}\ }\href {\doibase 10.1063/1.1700722} {\bibfield  {journal}
  {\bibinfo  {journal} {J. Chem. Phys.}\ }\textbf {\bibinfo {volume} {20}},\
  \bibinfo {pages} {1281--1295} (\bibinfo {year} {1952})}\BibitemShut {NoStop}%
\bibitem [{\citenamefont {Green}(1954)}]{Green1954}%
  \BibitemOpen
  \bibfield  {author} {\bibinfo {author} {\bibfnamefont {Melville~S.}\
  \bibnamefont {Green}},\ }\bibfield  {title} {\enquote {\bibinfo {title}
  {Markoff random processes and the statistical mechanics of time-dependent
  phenomena. ii. irreversible processes in fluids},}\ }\href {\doibase
  10.1063/1.1740082} {\bibfield  {journal} {\bibinfo  {journal} {J. Chem.
  Phys.}\ }\textbf {\bibinfo {volume} {22}},\ \bibinfo {pages} {398--413}
  (\bibinfo {year} {1954})}\BibitemShut {NoStop}%
\bibitem [{\citenamefont {Kubo}\ \emph {et~al.}(1957)\citenamefont {Kubo},
  \citenamefont {Yokota},\ and\ \citenamefont {Nakajima}}]{Kubo1957b}%
  \BibitemOpen
  \bibfield  {author} {\bibinfo {author} {\bibfnamefont {Ryogo}\ \bibnamefont
  {Kubo}}, \bibinfo {author} {\bibfnamefont {Mario}\ \bibnamefont {Yokota}}, \
  and\ \bibinfo {author} {\bibfnamefont {Sadao}\ \bibnamefont {Nakajima}},\
  }\bibfield  {title} {\enquote {\bibinfo {title} {Statistical-mechanical
  theory of irreversible processes. ii. response to thermal disturbance},}\
  }\href {\doibase 10.1143/JPSJ.12.1203} {\bibfield  {journal} {\bibinfo
  {journal} {J. Phys. Soc. Jpn.}\ }\textbf {\bibinfo {volume} {12}},\ \bibinfo
  {pages} {1203--1211} (\bibinfo {year} {1957})}\BibitemShut {NoStop}%
\bibitem [{\citenamefont {Kubo}(1957)}]{Kubo1957a}%
  \BibitemOpen
  \bibfield  {author} {\bibinfo {author} {\bibfnamefont {R}~\bibnamefont
  {Kubo}},\ }\bibfield  {title} {\enquote {\bibinfo {title}
  {Statistical-mechanical theory of irreversible processes. i. {General} theory
  and simple applications to magnetic and conduction problems},}\ }\href
  {\doibase 10.1143/JPSJ.12.570} {\bibfield  {journal} {\bibinfo  {journal} {J.
  Phys. Soc. Jpn.}\ }\textbf {\bibinfo {volume} {12}},\ \bibinfo {pages}
  {570--586} (\bibinfo {year} {1957})}\BibitemShut {NoStop}%
\bibitem [{\citenamefont {Stackhouse}\ \emph {et~al.}(2010)\citenamefont
  {Stackhouse}, \citenamefont {Stixrude},\ and\ \citenamefont
  {Karki}}]{Stackhouse2010b}%
  \BibitemOpen
  \bibfield  {author} {\bibinfo {author} {\bibfnamefont {Stephen}\ \bibnamefont
  {Stackhouse}}, \bibinfo {author} {\bibfnamefont {Lars}\ \bibnamefont
  {Stixrude}}, \ and\ \bibinfo {author} {\bibfnamefont {Bijaya~B.}\
  \bibnamefont {Karki}},\ }\bibfield  {title} {\enquote {\bibinfo {title}
  {Thermal conductivity of periclase ({MgO}) from first principles},}\ }\href
  {\doibase 10.1103/PhysRevLett.104.208501} {\bibfield  {journal} {\bibinfo
  {journal} {Phys. Rev. Lett.}\ }\textbf {\bibinfo {volume} {104}},\ \bibinfo
  {pages} {208501} (\bibinfo {year} {2010})}\BibitemShut {NoStop}%
\bibitem [{\citenamefont {Marcolongo}\ \emph {et~al.}(2016)\citenamefont
  {Marcolongo}, \citenamefont {Umari},\ and\ \citenamefont
  {Baroni}}]{Marcolongo2016}%
  \BibitemOpen
  \bibfield  {author} {\bibinfo {author} {\bibfnamefont {Aris}\ \bibnamefont
  {Marcolongo}}, \bibinfo {author} {\bibfnamefont {Paolo}\ \bibnamefont
  {Umari}}, \ and\ \bibinfo {author} {\bibfnamefont {Stefano}\ \bibnamefont
  {Baroni}},\ }\bibfield  {title} {\enquote {\bibinfo {title} {Microscopic
  theory and ab initio simulation of atomic heat transport},}\ }\href {\doibase
  10.1038/nphys3509} {\bibfield  {journal} {\bibinfo  {journal} {Nature Phys.}\
  }\textbf {\bibinfo {volume} {12}},\ \bibinfo {pages} {80--84} (\bibinfo
  {year} {2016})}\BibitemShut {NoStop}%
\bibitem [{\citenamefont {Ercole}\ \emph {et~al.}(2016)\citenamefont {Ercole},
  \citenamefont {Marcolongo}, \citenamefont {Umari},\ and\ \citenamefont
  {Baroni}}]{Ercole2016}%
  \BibitemOpen
  \bibfield  {author} {\bibinfo {author} {\bibfnamefont {Loris}\ \bibnamefont
  {Ercole}}, \bibinfo {author} {\bibfnamefont {Aris}\ \bibnamefont
  {Marcolongo}}, \bibinfo {author} {\bibfnamefont {Paolo}\ \bibnamefont
  {Umari}}, \ and\ \bibinfo {author} {\bibfnamefont {Stefano}\ \bibnamefont
  {Baroni}},\ }\bibfield  {title} {\enquote {\bibinfo {title} {Gauge invariance
  of thermal transport coefficients},}\ }\href {\doibase
  10.1007/s10909-016-1617-6} {\bibfield  {journal} {\bibinfo  {journal} {J. Low
  Temp. Phys.}\ }\textbf {\bibinfo {volume} {185}},\ \bibinfo {pages} {79--86}
  (\bibinfo {year} {2016})}\BibitemShut {NoStop}%
\bibitem [{\citenamefont {Baroni}\ \emph {et~al.}(2018)\citenamefont {Baroni},
  \citenamefont {Bertossa}, \citenamefont {Ercole}, \citenamefont {Grasselli},\
  and\ \citenamefont {Marcolongo}}]{Baroni2018}%
  \BibitemOpen
  \bibfield  {author} {\bibinfo {author} {\bibfnamefont {Stefano}\ \bibnamefont
  {Baroni}}, \bibinfo {author} {\bibfnamefont {Riccardo}\ \bibnamefont
  {Bertossa}}, \bibinfo {author} {\bibfnamefont {Loris}\ \bibnamefont
  {Ercole}}, \bibinfo {author} {\bibfnamefont {Federico}\ \bibnamefont
  {Grasselli}}, \ and\ \bibinfo {author} {\bibfnamefont {Aris}\ \bibnamefont
  {Marcolongo}},\ }\enquote {\bibinfo {title} {Heat transport in insulators
  from ab initio green-kubo theory},}\ in\ \href {\doibase
  10.1007/978-3-319-50257-1_12-1} {\emph {\bibinfo {booktitle} {Handbook of
  Materials Modeling: Applications: Current and Emerging Materials}}},\
  \bibinfo {editor} {edited by\ \bibinfo {editor} {\bibfnamefont {Wanda}\
  \bibnamefont {Andreoni}}\ and\ \bibinfo {editor} {\bibfnamefont {Sidney}\
  \bibnamefont {Yip}}}\ (\bibinfo  {publisher} {Springer International
  Publishing},\ \bibinfo {address} {Cham},\ \bibinfo {year} {2018})\ pp.\
  \bibinfo {pages} {1--36},\ \bibinfo {edition} {2nd}\ ed.,\ \Eprint
  {http://arxiv.org/abs/1802.08006} {arXiv:1802.08006 [cond-mat.stat-mech]}
  \BibitemShut {NoStop}%
\bibitem [{\citenamefont {Bertossa}\ \emph {et~al.}(2019)\citenamefont
  {Bertossa}, \citenamefont {Grasselli}, \citenamefont {Ercole},\ and\
  \citenamefont {Baroni}}]{Bertossa2019}%
  \BibitemOpen
  \bibfield  {author} {\bibinfo {author} {\bibfnamefont {Riccardo}\
  \bibnamefont {Bertossa}}, \bibinfo {author} {\bibfnamefont {Federico}\
  \bibnamefont {Grasselli}}, \bibinfo {author} {\bibfnamefont {Loris}\
  \bibnamefont {Ercole}}, \ and\ \bibinfo {author} {\bibfnamefont {Stefano}\
  \bibnamefont {Baroni}},\ }\bibfield  {title} {\enquote {\bibinfo {title}
  {Theory and numerical simulation of heat transport in multicomponent
  systems},}\ }\href {\doibase 10.1103/PhysRevLett.122.255901} {\bibfield
  {journal} {\bibinfo  {journal} {Phys. Rev. Lett.}\ }\textbf {\bibinfo
  {volume} {122}},\ \bibinfo {pages} {255901} (\bibinfo {year}
  {2019})}\BibitemShut {NoStop}%
\bibitem [{\citenamefont {Kadanoff}\ and\ \citenamefont
  {Martin}(1963)}]{Kadanoff1963}%
  \BibitemOpen
  \bibfield  {author} {\bibinfo {author} {\bibfnamefont {Leo~P}\ \bibnamefont
  {Kadanoff}}\ and\ \bibinfo {author} {\bibfnamefont {Paul~C}\ \bibnamefont
  {Martin}},\ }\bibfield  {title} {\enquote {\bibinfo {title} {{Hydrodynamic
  equations and correlation functions}},}\ }\href {\doibase
  10.1016/0003-4916(63)90078-2} {\bibfield  {journal} {\bibinfo  {journal}
  {Ann. Phys.}\ }\textbf {\bibinfo {volume} {24}},\ \bibinfo {pages} {419--469}
  (\bibinfo {year} {1963})}\BibitemShut {NoStop}%
\bibitem [{\citenamefont {Foster}(1975)}]{Foster1975}%
  \BibitemOpen
  \bibfield  {author} {\bibinfo {author} {\bibfnamefont {D.}~\bibnamefont
  {Foster}},\ }\href@noop {} {\emph {\bibinfo {title} {Hydrodynamic
  fluctuations, broken symmetry, and correlation functions}}}\ (\bibinfo
  {publisher} {Benjamin},\ \bibinfo {year} {1975})\BibitemShut {NoStop}%
\bibitem [{\citenamefont {Wiener}(1930)}]{Wiener1930}%
  \BibitemOpen
  \bibfield  {author} {\bibinfo {author} {\bibfnamefont {N.}~\bibnamefont
  {Wiener}},\ }\bibfield  {title} {\enquote {\bibinfo {title} {{Generalized
  harmonic analysis}},}\ }\href {\doibase 10.1007/BF02546511} {\bibfield
  {journal} {\bibinfo  {journal} {Acta Math.}\ }\textbf {\bibinfo {volume}
  {55}},\ \bibinfo {pages} {117--258} (\bibinfo {year} {1930})}\BibitemShut
  {NoStop}%
\bibitem [{\citenamefont {Khintchine}(1934)}]{Khintchine1934}%
  \BibitemOpen
  \bibfield  {author} {\bibinfo {author} {\bibfnamefont {A}~\bibnamefont
  {Khintchine}},\ }\bibfield  {title} {\enquote {\bibinfo {title}
  {{Korrelationstheorie der station{\"{a}}ren stochastischen Prozesse}},}\
  }\href {\doibase 10.1007/BF01449156} {\bibfield  {journal} {\bibinfo
  {journal} {Math. Ann.}\ }\textbf {\bibinfo {volume} {109}},\ \bibinfo {pages}
  {604--615} (\bibinfo {year} {1934})}\BibitemShut {NoStop}%
\bibitem [{\citenamefont {Irving}\ and\ \citenamefont
  {Kirkwood}(1950)}]{Irving1950}%
  \BibitemOpen
  \bibfield  {author} {\bibinfo {author} {\bibfnamefont {J~H}\ \bibnamefont
  {Irving}}\ and\ \bibinfo {author} {\bibfnamefont {J~G}\ \bibnamefont
  {Kirkwood}},\ }\bibfield  {title} {\enquote {\bibinfo {title} {{The
  statistical mechanical theory of transport processes. IV. The equations of
  hydrodynamics}},}\ }\href {\doibase 10.1063/1.1747782} {\bibfield  {journal}
  {\bibinfo  {journal} {J. Chem. Phys.}\ }\textbf {\bibinfo {volume} {18}},\
  \bibinfo {pages} {817} (\bibinfo {year} {1950})}\BibitemShut {NoStop}%
\bibitem [{\citenamefont {Groot}\ and\ \citenamefont
  {Mazur}(1984)}]{DeGroot1984}%
  \BibitemOpen
  \bibfield  {author} {\bibinfo {author} {\bibfnamefont {S.~R.~De}\
  \bibnamefont {Groot}}\ and\ \bibinfo {author} {\bibfnamefont
  {P.}~\bibnamefont {Mazur}},\ }\href@noop {} {\emph {\bibinfo {title}
  {Non-Equilibrium Thermodynamics}}}\ (\bibinfo  {publisher} {Dover
  Publications},\ \bibinfo {year} {1984})\BibitemShut {NoStop}%
\bibitem [{\citenamefont {Debenedetti}(1987)}]{Debenedetti1987}%
  \BibitemOpen
  \bibfield  {author} {\bibinfo {author} {\bibfnamefont {Pablo~G}\ \bibnamefont
  {Debenedetti}},\ }\bibfield  {title} {\enquote {\bibinfo {title}
  {{Fluctuation-based computer calculation of partial molar properties . I .
  Molecular dynamics simulation of constant volume fluctuations Molecular
  dynamics simulation of constant volume fluctuations}},}\ }\href {\doibase
  10.1063/1.452362} {\bibfield  {journal} {\bibinfo  {journal} {J. Chem.
  Phys.}\ }\textbf {\bibinfo {volume} {86}},\ \bibinfo {pages} {7126} (\bibinfo
  {year} {1987})}\BibitemShut {NoStop}%
\bibitem [{\citenamefont {Vogelsang}\ and\ \citenamefont
  {Hoheisel}(1987)}]{Vogelsang1987}%
  \BibitemOpen
  \bibfield  {author} {\bibinfo {author} {\bibfnamefont {R}~\bibnamefont
  {Vogelsang}}\ and\ \bibinfo {author} {\bibfnamefont {C}~\bibnamefont
  {Hoheisel}},\ }\bibfield  {title} {\enquote {\bibinfo {title} {{Thermal
  conductivity of a binary-liquid mixture studied by molecular dynamics with
  use of Lennard- Jones potentials}},}\ }\href@noop {} {\bibfield  {journal}
  {\bibinfo  {journal} {Phys. Rev. A.}\ }\textbf {\bibinfo {volume} {35}},\
  \bibinfo {pages} {3487--3491} (\bibinfo {year} {1987})}\BibitemShut {NoStop}%
\bibitem [{\citenamefont {Sindzingre}\ \emph {et~al.}(1989)\citenamefont
  {Sindzingre}, \citenamefont {Massobrio},\ and\ \citenamefont
  {Ciccotti}}]{Sindzingre1989}%
  \BibitemOpen
  \bibfield  {author} {\bibinfo {author} {\bibfnamefont {P}~\bibnamefont
  {Sindzingre}}, \bibinfo {author} {\bibfnamefont {C}~\bibnamefont
  {Massobrio}}, \ and\ \bibinfo {author} {\bibfnamefont {G}~\bibnamefont
  {Ciccotti}},\ }\bibfield  {title} {\enquote {\bibinfo {title} {{Calculation
  of partial enthalpies of an Argon-Kripton mixture by NPT molecular
  dynamics}},}\ }\href@noop {} {\bibfield  {journal} {\bibinfo  {journal} {Chem
  Phys.}\ }\textbf {\bibinfo {volume} {129}},\ \bibinfo {pages} {213--224}
  (\bibinfo {year} {1989})}\BibitemShut {NoStop}%
\bibitem [{\citenamefont {Onsager}(1931{\natexlab{a}})}]{Onsager1931a}%
  \BibitemOpen
  \bibfield  {author} {\bibinfo {author} {\bibfnamefont {Lars}\ \bibnamefont
  {Onsager}},\ }\bibfield  {title} {\enquote {\bibinfo {title} {Reciprocal
  relations in irreversible processes. i.}}\ }\href {\doibase
  10.1103/PhysRev.37.405} {\bibfield  {journal} {\bibinfo  {journal} {Phys.
  Rev.}\ }\textbf {\bibinfo {volume} {37}},\ \bibinfo {pages} {405--426}
  (\bibinfo {year} {1931}{\natexlab{a}})}\BibitemShut {NoStop}%
\bibitem [{\citenamefont {Onsager}(1931{\natexlab{b}})}]{Onsager1931b}%
  \BibitemOpen
  \bibfield  {author} {\bibinfo {author} {\bibfnamefont {L}~\bibnamefont
  {Onsager}},\ }\bibfield  {title} {\enquote {\bibinfo {title} {Reciprocal
  relations in irreversible processes. ii.}}\ }\href {\doibase
  10.1103/PhysRev.38.2265} {\bibfield  {journal} {\bibinfo  {journal} {Phys.
  Rev.}\ }\textbf {\bibinfo {volume} {38}},\ \bibinfo {pages} {2265} (\bibinfo
  {year} {1931}{\natexlab{b}})}\BibitemShut {NoStop}%
\bibitem [{\citenamefont {Galamba}\ \emph {et~al.}(2007)\citenamefont
  {Galamba}, \citenamefont {{Nieto de Castro}},\ and\ \citenamefont
  {Ely}}]{Galamba2007}%
  \BibitemOpen
  \bibfield  {author} {\bibinfo {author} {\bibfnamefont {N}~\bibnamefont
  {Galamba}}, \bibinfo {author} {\bibfnamefont {C~a}\ \bibnamefont {{Nieto de
  Castro}}}, \ and\ \bibinfo {author} {\bibfnamefont {James~F}\ \bibnamefont
  {Ely}},\ }\bibfield  {title} {\enquote {\bibinfo {title} {{Equilibrium and
  nonequilibrium molecular dynamics simulations of the thermal conductivity of
  molten alkali halides.}}}\ }\href {\doibase 10.1063/1.2734965} {\bibfield
  {journal} {\bibinfo  {journal} {J. Chem. Phys.}\ }\textbf {\bibinfo {volume}
  {126}},\ \bibinfo {pages} {204511} (\bibinfo {year} {2007})}\BibitemShut
  {NoStop}%
\bibitem [{\citenamefont {Ohtori}\ \emph {et~al.}(2009)\citenamefont {Ohtori},
  \citenamefont {Salanne},\ and\ \citenamefont {Madden}}]{Ohtori2009b}%
  \BibitemOpen
  \bibfield  {author} {\bibinfo {author} {\bibfnamefont {Norikazu}\
  \bibnamefont {Ohtori}}, \bibinfo {author} {\bibfnamefont {Mathieu}\
  \bibnamefont {Salanne}}, \ and\ \bibinfo {author} {\bibfnamefont {Paul~A.}\
  \bibnamefont {Madden}},\ }\bibfield  {title} {\enquote {\bibinfo {title}
  {Calculations of the thermal conductivities of ionic materials by simulation
  with polarizable interaction potentials},}\ }\href {\doibase
  10.1063/1.3086856} {\bibfield  {journal} {\bibinfo  {journal} {J. Chem.
  Phys.}\ }\textbf {\bibinfo {volume} {130}},\ \bibinfo {pages} {104507}
  (\bibinfo {year} {2009})}\BibitemShut {NoStop}%
\bibitem [{\citenamefont {Salanne}\ \emph {et~al.}(2011)\citenamefont
  {Salanne}, \citenamefont {Marrocchelli}, \citenamefont {Merlet},
  \citenamefont {Ohtori},\ and\ \citenamefont {Madden}}]{Salanne2011}%
  \BibitemOpen
  \bibfield  {author} {\bibinfo {author} {\bibfnamefont {Mathieu}\ \bibnamefont
  {Salanne}}, \bibinfo {author} {\bibfnamefont {Dario}\ \bibnamefont
  {Marrocchelli}}, \bibinfo {author} {\bibfnamefont {C{\'{e}}line}\
  \bibnamefont {Merlet}}, \bibinfo {author} {\bibfnamefont {Norikazu}\
  \bibnamefont {Ohtori}}, \ and\ \bibinfo {author} {\bibfnamefont {Paul~A}\
  \bibnamefont {Madden}},\ }\bibfield  {title} {\enquote {\bibinfo {title}
  {{Thermal conductivity of ionic systems from equilibrium molecular
  dynamics}},}\ }\href {\doibase 10.1088/0953-8984/23/10/102101} {\bibfield
  {journal} {\bibinfo  {journal} {J. Phys. Condens. Matter}\ }\textbf {\bibinfo
  {volume} {23}},\ \bibinfo {pages} {102101} (\bibinfo {year}
  {2011})}\BibitemShut {NoStop}%
\bibitem [{\citenamefont {Bonella}\ \emph {et~al.}(2017)\citenamefont
  {Bonella}, \citenamefont {Ferrario},\ and\ \citenamefont
  {Ciccotti}}]{Bonella2017}%
  \BibitemOpen
  \bibfield  {author} {\bibinfo {author} {\bibfnamefont {Sara}\ \bibnamefont
  {Bonella}}, \bibinfo {author} {\bibfnamefont {Mauro}\ \bibnamefont
  {Ferrario}}, \ and\ \bibinfo {author} {\bibfnamefont {Giovanni}\ \bibnamefont
  {Ciccotti}},\ }\bibfield  {title} {\enquote {\bibinfo {title} {Thermal
  diffusion in binary mixtures: Transient behavior and transport coefficients
  from equilibrium and nonequilibrium molecular dynamics},}\ }\href {\doibase
  10.1021/acs.langmuir.7b02565} {\bibfield  {journal} {\bibinfo  {journal}
  {Langmuir}\ }\textbf {\bibinfo {volume} {33}},\ \bibinfo {pages}
  {11281--11290} (\bibinfo {year} {2017})}\BibitemShut {NoStop}%
\bibitem [{\citenamefont {Nagar}\ and\ \citenamefont
  {Gupta}(2011)}]{Nagar2011}%
  \BibitemOpen
  \bibfield  {author} {\bibinfo {author} {\bibfnamefont {Daya~K.}\ \bibnamefont
  {Nagar}}\ and\ \bibinfo {author} {\bibfnamefont {Arjun~K.}\ \bibnamefont
  {Gupta}},\ }\bibfield  {title} {\enquote {\bibinfo {title} {{Expectations of
  functions of complex wishart matrix}},}\ }\href {\doibase
  10.1007/s10440-010-9599-x} {\bibfield  {journal} {\bibinfo  {journal} {Acta
  Appl. Math.}\ }\textbf {\bibinfo {volume} {113}},\ \bibinfo {pages}
  {265--288} (\bibinfo {year} {2011})}\BibitemShut {NoStop}%
\bibitem [{\citenamefont {Childers}\ \emph {et~al.}(1977)\citenamefont
  {Childers}, \citenamefont {Skinner},\ and\ \citenamefont
  {Kemerait}}]{Childers1977}%
  \BibitemOpen
  \bibfield  {author} {\bibinfo {author} {\bibfnamefont {D.~G.}\ \bibnamefont
  {Childers}}, \bibinfo {author} {\bibfnamefont {D.~P.}\ \bibnamefont
  {Skinner}}, \ and\ \bibinfo {author} {\bibfnamefont {R.~C.}\ \bibnamefont
  {Kemerait}},\ }\bibfield  {title} {\enquote {\bibinfo {title} {The cepstrum:
  A guide to processing},}\ }\href {\doibase 10.1109/PROC.1977.10747}
  {\bibfield  {journal} {\bibinfo  {journal} {Proceedings of the IEEE}\
  }\textbf {\bibinfo {volume} {65}},\ \bibinfo {pages} {1428--1443} (\bibinfo
  {year} {1977})}\BibitemShut {NoStop}%
\bibitem [{\citenamefont {Car}\ and\ \citenamefont
  {Parrinello}(1985)}]{Car1985}%
  \BibitemOpen
  \bibfield  {author} {\bibinfo {author} {\bibfnamefont {R.}~\bibnamefont
  {Car}}\ and\ \bibinfo {author} {\bibfnamefont {M.}~\bibnamefont
  {Parrinello}},\ }\bibfield  {title} {\enquote {\bibinfo {title} {Unified
  approach for molecular dynamics and density-functional theory},}\ }\href
  {\doibase 10.1103/PhysRevLett.55.2471} {\bibfield  {journal} {\bibinfo
  {journal} {Phys. Rev. Lett.}\ }\textbf {\bibinfo {volume} {55}},\ \bibinfo
  {pages} {2471--2474} (\bibinfo {year} {1985})}\BibitemShut {NoStop}%
\bibitem [{\citenamefont {Giannozzi}\ \emph {et~al.}(2009)\citenamefont
  {Giannozzi}, \citenamefont {Baroni}, \citenamefont {Bonini}, \citenamefont
  {Calandra}, \citenamefont {Car}, \citenamefont {Cavazzoni}, \citenamefont
  {Ceresoli}, \citenamefont {Chiarotti}, \citenamefont {Cococcioni},
  \citenamefont {Dabo}, \citenamefont {Corso}, \citenamefont {Gironcoli},
  \citenamefont {Fabris}, \citenamefont {Fratesi}, \citenamefont {Gebauer},
  \citenamefont {Gerstmann}, \citenamefont {Gougoussis}, \citenamefont
  {Kokalj}, \citenamefont {Lazzeri}, \citenamefont {Martin-Samos},
  \citenamefont {Marzari}, \citenamefont {Mauri}, \citenamefont {Mazzarello},
  \citenamefont {Paolini}, \citenamefont {Pasquarello}, \citenamefont
  {Paulatto}, \citenamefont {Sbraccia}, \citenamefont {Scandolo}, \citenamefont
  {Sclauzero}, \citenamefont {Seitsonen}, \citenamefont {Smogunov},
  \citenamefont {Umari},\ and\ \citenamefont {Wentzcovitch}}]{Giannozzi2009}%
  \BibitemOpen
  \bibfield  {author} {\bibinfo {author} {\bibfnamefont {P}~\bibnamefont
  {Giannozzi}}, \bibinfo {author} {\bibfnamefont {S}~\bibnamefont {Baroni}},
  \bibinfo {author} {\bibfnamefont {N}~\bibnamefont {Bonini}}, \bibinfo
  {author} {\bibfnamefont {M}~\bibnamefont {Calandra}}, \bibinfo {author}
  {\bibfnamefont {R}~\bibnamefont {Car}}, \bibinfo {author} {\bibfnamefont
  {C}~\bibnamefont {Cavazzoni}}, \bibinfo {author} {\bibfnamefont
  {D}~\bibnamefont {Ceresoli}}, \bibinfo {author} {\bibfnamefont {G~L}\
  \bibnamefont {Chiarotti}}, \bibinfo {author} {\bibfnamefont {M}~\bibnamefont
  {Cococcioni}}, \bibinfo {author} {\bibfnamefont {I}~\bibnamefont {Dabo}},
  \bibinfo {author} {\bibfnamefont {A~D}\ \bibnamefont {Corso}}, \bibinfo
  {author} {\bibfnamefont {Sd}~\bibnamefont {Gironcoli}}, \bibinfo {author}
  {\bibfnamefont {S}~\bibnamefont {Fabris}}, \bibinfo {author} {\bibfnamefont
  {G}~\bibnamefont {Fratesi}}, \bibinfo {author} {\bibfnamefont
  {R}~\bibnamefont {Gebauer}}, \bibinfo {author} {\bibfnamefont
  {U}~\bibnamefont {Gerstmann}}, \bibinfo {author} {\bibfnamefont
  {C}~\bibnamefont {Gougoussis}}, \bibinfo {author} {\bibfnamefont
  {A}~\bibnamefont {Kokalj}}, \bibinfo {author} {\bibfnamefont {M}~\bibnamefont
  {Lazzeri}}, \bibinfo {author} {\bibfnamefont {L}~\bibnamefont
  {Martin-Samos}}, \bibinfo {author} {\bibfnamefont {N}~\bibnamefont
  {Marzari}}, \bibinfo {author} {\bibfnamefont {F}~\bibnamefont {Mauri}},
  \bibinfo {author} {\bibfnamefont {R}~\bibnamefont {Mazzarello}}, \bibinfo
  {author} {\bibfnamefont {S}~\bibnamefont {Paolini}}, \bibinfo {author}
  {\bibfnamefont {A}~\bibnamefont {Pasquarello}}, \bibinfo {author}
  {\bibfnamefont {L}~\bibnamefont {Paulatto}}, \bibinfo {author} {\bibfnamefont
  {C}~\bibnamefont {Sbraccia}}, \bibinfo {author} {\bibfnamefont
  {S}~\bibnamefont {Scandolo}}, \bibinfo {author} {\bibfnamefont
  {G}~\bibnamefont {Sclauzero}}, \bibinfo {author} {\bibfnamefont {A~P}\
  \bibnamefont {Seitsonen}}, \bibinfo {author} {\bibfnamefont {A}~\bibnamefont
  {Smogunov}}, \bibinfo {author} {\bibfnamefont {P}~\bibnamefont {Umari}}, \
  and\ \bibinfo {author} {\bibfnamefont {R~M}\ \bibnamefont {Wentzcovitch}},\
  }\bibfield  {title} {\enquote {\bibinfo {title} {{QUANTUM ESPRESSO}: a
  modular and open-source software project for quantum simulations of
  materials},}\ }\href {\doibase 10.1088/0953-8984/21/39/395502} {\bibfield
  {journal} {\bibinfo  {journal} {J. Phys. Condens. Matter}\ }\textbf {\bibinfo
  {volume} {21}},\ \bibinfo {pages} {395502 (19pp)} (\bibinfo {year}
  {2009})}\BibitemShut {NoStop}%
\bibitem [{\citenamefont {Giannozzi}\ \emph {et~al.}(2017)\citenamefont
  {Giannozzi}, \citenamefont {Andreussi}, \citenamefont {Brumme}, \citenamefont
  {Bunau}, \citenamefont {Nardelli}, \citenamefont {Calandra}, \citenamefont
  {Car}, \citenamefont {Cavazzoni}, \citenamefont {Ceresoli}, \citenamefont
  {Cococcioni}, \citenamefont {Colonna}, \citenamefont {Carnimeo},
  \citenamefont {Corso}, \citenamefont {de~Gironcoli}, \citenamefont {Delugas},
  \citenamefont {Jr}, \citenamefont {Ferretti}, \citenamefont {Floris},
  \citenamefont {Fratesi}, \citenamefont {Fugallo}, \citenamefont {Gebauer},
  \citenamefont {Gerstmann}, \citenamefont {Giustino}, \citenamefont {Gorni},
  \citenamefont {Jia}, \citenamefont {Kawamura}, \citenamefont {Ko},
  \citenamefont {Kokalj}, \citenamefont {Küçükbenli}, \citenamefont
  {Lazzeri}, \citenamefont {Marsili}, \citenamefont {Marzari}, \citenamefont
  {Mauri}, \citenamefont {Nguyen}, \citenamefont {Nguyen}, \citenamefont {de-la
  Roza}, \citenamefont {Paulatto}, \citenamefont {Poncé}, \citenamefont
  {Rocca}, \citenamefont {Sabatini}, \citenamefont {Santra}, \citenamefont
  {Schlipf}, \citenamefont {Seitsonen}, \citenamefont {Smogunov}, \citenamefont
  {Timrov}, \citenamefont {Thonhauser}, \citenamefont {Umari}, \citenamefont
  {Vast}, \citenamefont {Wu},\ and\ \citenamefont {Baroni}}]{Giannozzi2017}%
  \BibitemOpen
  \bibfield  {author} {\bibinfo {author} {\bibfnamefont {P}~\bibnamefont
  {Giannozzi}}, \bibinfo {author} {\bibfnamefont {O}~\bibnamefont {Andreussi}},
  \bibinfo {author} {\bibfnamefont {T}~\bibnamefont {Brumme}}, \bibinfo
  {author} {\bibfnamefont {O}~\bibnamefont {Bunau}}, \bibinfo {author}
  {\bibfnamefont {M~Buongiorno}\ \bibnamefont {Nardelli}}, \bibinfo {author}
  {\bibfnamefont {M}~\bibnamefont {Calandra}}, \bibinfo {author} {\bibfnamefont
  {R}~\bibnamefont {Car}}, \bibinfo {author} {\bibfnamefont {C}~\bibnamefont
  {Cavazzoni}}, \bibinfo {author} {\bibfnamefont {D}~\bibnamefont {Ceresoli}},
  \bibinfo {author} {\bibfnamefont {M}~\bibnamefont {Cococcioni}}, \bibinfo
  {author} {\bibfnamefont {N}~\bibnamefont {Colonna}}, \bibinfo {author}
  {\bibfnamefont {I}~\bibnamefont {Carnimeo}}, \bibinfo {author} {\bibfnamefont
  {A~Dal}\ \bibnamefont {Corso}}, \bibinfo {author} {\bibfnamefont
  {S}~\bibnamefont {de~Gironcoli}}, \bibinfo {author} {\bibfnamefont
  {P}~\bibnamefont {Delugas}}, \bibinfo {author} {\bibfnamefont {R~A~DiStasio}\
  \bibnamefont {Jr}}, \bibinfo {author} {\bibfnamefont {A}~\bibnamefont
  {Ferretti}}, \bibinfo {author} {\bibfnamefont {A}~\bibnamefont {Floris}},
  \bibinfo {author} {\bibfnamefont {G}~\bibnamefont {Fratesi}}, \bibinfo
  {author} {\bibfnamefont {G}~\bibnamefont {Fugallo}}, \bibinfo {author}
  {\bibfnamefont {R}~\bibnamefont {Gebauer}}, \bibinfo {author} {\bibfnamefont
  {U}~\bibnamefont {Gerstmann}}, \bibinfo {author} {\bibfnamefont
  {F}~\bibnamefont {Giustino}}, \bibinfo {author} {\bibfnamefont
  {T}~\bibnamefont {Gorni}}, \bibinfo {author} {\bibfnamefont {J}~\bibnamefont
  {Jia}}, \bibinfo {author} {\bibfnamefont {M}~\bibnamefont {Kawamura}},
  \bibinfo {author} {\bibfnamefont {H-Y}\ \bibnamefont {Ko}}, \bibinfo {author}
  {\bibfnamefont {A}~\bibnamefont {Kokalj}}, \bibinfo {author} {\bibfnamefont
  {E}~\bibnamefont {Küçükbenli}}, \bibinfo {author} {\bibfnamefont
  {M}~\bibnamefont {Lazzeri}}, \bibinfo {author} {\bibfnamefont
  {M}~\bibnamefont {Marsili}}, \bibinfo {author} {\bibfnamefont
  {N}~\bibnamefont {Marzari}}, \bibinfo {author} {\bibfnamefont
  {F}~\bibnamefont {Mauri}}, \bibinfo {author} {\bibfnamefont {N~L}\
  \bibnamefont {Nguyen}}, \bibinfo {author} {\bibfnamefont {H-V}\ \bibnamefont
  {Nguyen}}, \bibinfo {author} {\bibfnamefont {A~Otero}\ \bibnamefont {de-la
  Roza}}, \bibinfo {author} {\bibfnamefont {L}~\bibnamefont {Paulatto}},
  \bibinfo {author} {\bibfnamefont {S}~\bibnamefont {Poncé}}, \bibinfo
  {author} {\bibfnamefont {D}~\bibnamefont {Rocca}}, \bibinfo {author}
  {\bibfnamefont {R}~\bibnamefont {Sabatini}}, \bibinfo {author} {\bibfnamefont
  {B}~\bibnamefont {Santra}}, \bibinfo {author} {\bibfnamefont {M}~\bibnamefont
  {Schlipf}}, \bibinfo {author} {\bibfnamefont {A~P}\ \bibnamefont
  {Seitsonen}}, \bibinfo {author} {\bibfnamefont {A}~\bibnamefont {Smogunov}},
  \bibinfo {author} {\bibfnamefont {I}~\bibnamefont {Timrov}}, \bibinfo
  {author} {\bibfnamefont {T}~\bibnamefont {Thonhauser}}, \bibinfo {author}
  {\bibfnamefont {P}~\bibnamefont {Umari}}, \bibinfo {author} {\bibfnamefont
  {N}~\bibnamefont {Vast}}, \bibinfo {author} {\bibfnamefont {X}~\bibnamefont
  {Wu}}, \ and\ \bibinfo {author} {\bibfnamefont {S}~\bibnamefont {Baroni}},\
  }\bibfield  {title} {\enquote {\bibinfo {title} {Advanced capabilities for
  materials modelling with q uantum espresso},}\ }\href
  {http://stacks.iop.org/0953-8984/29/i=46/a=465901} {\bibfield  {journal}
  {\bibinfo  {journal} {Journal of Physics: Condensed Matter}\ }\textbf
  {\bibinfo {volume} {29}},\ \bibinfo {pages} {465901} (\bibinfo {year}
  {2017})}\BibitemShut {NoStop}%
\bibitem [{\citenamefont {Redmer}\ \emph {et~al.}(2011)\citenamefont {Redmer},
  \citenamefont {Mattsson}, \citenamefont {Nettelmann},\ and\ \citenamefont
  {French}}]{Redmer2011}%
  \BibitemOpen
  \bibfield  {author} {\bibinfo {author} {\bibfnamefont {Ronald}\ \bibnamefont
  {Redmer}}, \bibinfo {author} {\bibfnamefont {Thomas~R}\ \bibnamefont
  {Mattsson}}, \bibinfo {author} {\bibfnamefont {Nadine}\ \bibnamefont
  {Nettelmann}}, \ and\ \bibinfo {author} {\bibfnamefont {Martin}\ \bibnamefont
  {French}},\ }\bibfield  {title} {\enquote {\bibinfo {title} {The phase
  diagram of water and the magnetic fields of uranus and neptune},}\
  }\href@noop {} {\bibfield  {journal} {\bibinfo  {journal} {Icarus}\ }\textbf
  {\bibinfo {volume} {211}},\ \bibinfo {pages} {798--803} (\bibinfo {year}
  {2011})}\BibitemShut {NoStop}%
\bibitem [{\citenamefont {Millot}\ \emph {et~al.}(2019)\citenamefont {Millot},
  \citenamefont {Coppari}, \citenamefont {Rygg}, \citenamefont {Barrios},
  \citenamefont {Hamel}, \citenamefont {Swift},\ and\ \citenamefont
  {Eggert}}]{Millot2019}%
  \BibitemOpen
  \bibfield  {author} {\bibinfo {author} {\bibfnamefont {Marius}\ \bibnamefont
  {Millot}}, \bibinfo {author} {\bibfnamefont {Federica}\ \bibnamefont
  {Coppari}}, \bibinfo {author} {\bibfnamefont {J~Ryan}\ \bibnamefont {Rygg}},
  \bibinfo {author} {\bibfnamefont {Antonio~Correa}\ \bibnamefont {Barrios}},
  \bibinfo {author} {\bibfnamefont {Sebastien}\ \bibnamefont {Hamel}}, \bibinfo
  {author} {\bibfnamefont {Damian~C}\ \bibnamefont {Swift}}, \ and\ \bibinfo
  {author} {\bibfnamefont {Jon~H}\ \bibnamefont {Eggert}},\ }\bibfield  {title}
  {\enquote {\bibinfo {title} {Nanosecond x-ray diffraction of shock-compressed
  superionic water ice},}\ }\href@noop {} {\bibfield  {journal} {\bibinfo
  {journal} {Nature}\ }\textbf {\bibinfo {volume} {569}},\ \bibinfo {pages}
  {251--255} (\bibinfo {year} {2019})}\BibitemShut {NoStop}%
\bibitem [{\citenamefont {French}\ \emph {et~al.}(2010)\citenamefont {French},
  \citenamefont {Mattsson},\ and\ \citenamefont {Redmer}}]{French2010}%
  \BibitemOpen
  \bibfield  {author} {\bibinfo {author} {\bibfnamefont {Martin}\ \bibnamefont
  {French}}, \bibinfo {author} {\bibfnamefont {Thomas~R.}\ \bibnamefont
  {Mattsson}}, \ and\ \bibinfo {author} {\bibfnamefont {Ronald}\ \bibnamefont
  {Redmer}},\ }\bibfield  {title} {\enquote {\bibinfo {title} {Diffusion and
  electrical conductivity in water at ultrahigh pressures},}\ }\href {\doibase
  10.1103/PhysRevB.82.174108} {\bibfield  {journal} {\bibinfo  {journal} {Phys.
  Rev. B}\ }\textbf {\bibinfo {volume} {82}},\ \bibinfo {pages} {174108}
  (\bibinfo {year} {2010})}\BibitemShut {NoStop}%
\bibitem [{Sup()}]{SupplMat}%
  \BibitemOpen
  \href@noop {} {}\bibinfo {note} {See Supplementary Information, which contains Refs.
  \cite{Car1985,Giannozzi2009,Giannozzi2017,Vanderbilt85,Schlipf2015,Sun2015,Perdew1996,SunThesis2019,Nose1984,Hoover1985,Martyna:1992gy,Parrinello1980,Parrinello1981,Bernasconi1995,Zhang2018,Puligheddu2020,Marcolongo2016,Grasselli2019,Baroni2001,Ercole2017,thermocepstrum,Bertossa2019,Akaike1973,Akaike1974,MovingAverage,Goldsby2001}}\BibitemShut
  {NoStop}%
\bibitem [{\citenamefont {Marcolongo}\ \emph {et~al.}(2020)\citenamefont
  {Marcolongo}, \citenamefont {Ercole},\ and\ \citenamefont
  {Baroni}}]{antani2020}%
  \BibitemOpen
  \bibfield  {author} {\bibinfo {author} {\bibfnamefont {A}~\bibnamefont
  {Marcolongo}}, \bibinfo {author} {\bibfnamefont {L}~\bibnamefont {Ercole}}, \
  and\ \bibinfo {author} {\bibfnamefont {S}~\bibnamefont {Baroni}},\
  }\href@noop {} {\enquote {\bibinfo {title} {{Gauge fixing for thermal
  transport simulations}},}\ } (\bibinfo {year} {2020}),\ \bibinfo {note} {{J}.
  Chem. Theory Comput. in press},\ \Eprint {http://arxiv.org/abs/1909.13580}
  {arXiv:1909.13580 [physics.comp-ph]} \BibitemShut {NoStop}%
\bibitem [{\citenamefont {Ercole}\ and\ \citenamefont
  {Bertossa}(2017--2018)}]{thermocepstrum}%
  \BibitemOpen
  \bibfield  {author} {\bibinfo {author} {\bibfnamefont {Loris}\ \bibnamefont
  {Ercole}}\ and\ \bibinfo {author} {\bibfnamefont {Riccardo}\ \bibnamefont
  {Bertossa}},\ }\href@noop {} {\enquote {\bibinfo {title}
  {\texttt{ThermoCepstrum}: a code to estimate transport coefficients from the
  cepstral analysis of a multi-variate current stationary time series},}\
  }\bibinfo {howpublished}
  {\url{https://github.com/lorisercole/thermocepstrum}} (\bibinfo {year}
  {2017--2018})\BibitemShut {NoStop}%
\bibitem [{\citenamefont {Grasselli}\ and\ \citenamefont
  {Baroni}(2019)}]{Grasselli2019}%
  \BibitemOpen
  \bibfield  {author} {\bibinfo {author} {\bibfnamefont {Federico}\
  \bibnamefont {Grasselli}}\ and\ \bibinfo {author} {\bibfnamefont {Stefano}\
  \bibnamefont {Baroni}},\ }\bibfield  {title} {\enquote {\bibinfo {title}
  {Topological quantisation and gauge-invariance of charge transport in liquid
  insulators},}\ }\href {\doibase 10.1038/s41567-019-0562-0} {\bibfield
  {journal} {\bibinfo  {journal} {Nature Physics}\ }\textbf {\bibinfo {volume}
  {15}},\ \bibinfo {pages} {967--972} (\bibinfo {year} {2019})}\BibitemShut
  {NoStop}%
\bibitem [{\citenamefont {Marcolongo}\ and\ \citenamefont
  {Marzari}(2017)}]{Marcolongo2017}%
  \BibitemOpen
  \bibfield  {author} {\bibinfo {author} {\bibfnamefont {Aris}\ \bibnamefont
  {Marcolongo}}\ and\ \bibinfo {author} {\bibfnamefont {Nicola}\ \bibnamefont
  {Marzari}},\ }\bibfield  {title} {\enquote {\bibinfo {title} {Ionic
  correlations and failure of nernst-einstein relation in solid-state
  electrolytes},}\ }\href {\doibase 10.1103/PhysRevMaterials.1.025402}
  {\bibfield  {journal} {\bibinfo  {journal} {Phys. Rev. Materials}\ }\textbf
  {\bibinfo {volume} {1}},\ \bibinfo {pages} {025402} (\bibinfo {year}
  {2017})}\BibitemShut {NoStop}%
\bibitem [{\citenamefont {France-Lanord}\ and\ \citenamefont
  {Grossman}(2019)}]{France-Lanord2019}%
  \BibitemOpen
  \bibfield  {author} {\bibinfo {author} {\bibfnamefont {Arthur}\ \bibnamefont
  {France-Lanord}}\ and\ \bibinfo {author} {\bibfnamefont {Jeffrey~C.}\
  \bibnamefont {Grossman}},\ }\bibfield  {title} {\enquote {\bibinfo {title}
  {{Correlations from Ion Pairing and the Nernst-Einstein Equation}},}\ }\href
  {\doibase 10.1103/PhysRevLett.122.136001} {\bibfield  {journal} {\bibinfo
  {journal} {Physical Review Letters}\ }\textbf {\bibinfo {volume} {122}},\
  \bibinfo {pages} {1--6} (\bibinfo {year} {2019})},\ \Eprint
  {http://arxiv.org/abs/1812.04772} {arXiv:1812.04772} \BibitemShut {NoStop}%
\bibitem [{\citenamefont {Wilson}\ \emph {et~al.}(2013)\citenamefont {Wilson},
  \citenamefont {Wong},\ and\ \citenamefont {Militzer}}]{Wilson2013}%
  \BibitemOpen
  \bibfield  {author} {\bibinfo {author} {\bibfnamefont {Hugh~F.}\ \bibnamefont
  {Wilson}}, \bibinfo {author} {\bibfnamefont {Michael~L.}\ \bibnamefont
  {Wong}}, \ and\ \bibinfo {author} {\bibfnamefont {Burkhard}\ \bibnamefont
  {Militzer}},\ }\bibfield  {title} {\enquote {\bibinfo {title} {Superionic to
  superionic phase change in water: Consequences for the interiors of uranus
  and neptune},}\ }\href {\doibase 10.1103/PhysRevLett.110.151102} {\bibfield
  {journal} {\bibinfo  {journal} {Phys. Rev. Lett.}\ }\textbf {\bibinfo
  {volume} {110}},\ \bibinfo {pages} {151102} (\bibinfo {year}
  {2013})}\BibitemShut {NoStop}%
\bibitem [{\citenamefont {He}\ \emph {et~al.}(2017)\citenamefont {He},
  \citenamefont {Zhu},\ and\ \citenamefont {Mo}}]{He2017}%
  \BibitemOpen
  \bibfield  {author} {\bibinfo {author} {\bibfnamefont {Xingfeng}\
  \bibnamefont {He}}, \bibinfo {author} {\bibfnamefont {Yizhou}\ \bibnamefont
  {Zhu}}, \ and\ \bibinfo {author} {\bibfnamefont {Yifei}\ \bibnamefont {Mo}},\
  }\bibfield  {title} {\enquote {\bibinfo {title} {Origin of fast ion diffusion
  in super-ionic conductors},}\ }\href@noop {} {\bibfield  {journal} {\bibinfo
  {journal} {Nature communications}\ }\textbf {\bibinfo {volume} {8}},\
  \bibinfo {pages} {1--7} (\bibinfo {year} {2017})}\BibitemShut {NoStop}%
\bibitem [{\citenamefont {Podolak}\ \emph {et~al.}(2019)\citenamefont
  {Podolak}, \citenamefont {Helled},\ and\ \citenamefont
  {Schubert}}]{Podolak2019}%
  \BibitemOpen
  \bibfield  {author} {\bibinfo {author} {\bibfnamefont {Morris}\ \bibnamefont
  {Podolak}}, \bibinfo {author} {\bibfnamefont {Ravit}\ \bibnamefont {Helled}},
  \ and\ \bibinfo {author} {\bibfnamefont {Gerald}\ \bibnamefont {Schubert}},\
  }\bibfield  {title} {\enquote {\bibinfo {title} {Effect of non-adiabatic
  thermal profiles on the inferred compositions of uranus and neptune},}\
  }\href@noop {} {\bibfield  {journal} {\bibinfo  {journal} {Monthly Notices of
  the Royal Astronomical Society}\ }\textbf {\bibinfo {volume} {487}},\
  \bibinfo {pages} {2653--2664} (\bibinfo {year} {2019})}\BibitemShut {NoStop}%
\bibitem [{\citenamefont {French}(2019)}]{French2019}%
  \BibitemOpen
  \bibfield  {author} {\bibinfo {author} {\bibfnamefont {Martin}\ \bibnamefont
  {French}},\ }\bibfield  {title} {\enquote {\bibinfo {title} {Thermal
  conductivity of dissociating water—an ab initio study},}\ }\href@noop {}
  {\bibfield  {journal} {\bibinfo  {journal} {New Journal of Physics}\ }\textbf
  {\bibinfo {volume} {21}},\ \bibinfo {pages} {023007} (\bibinfo {year}
  {2019})}\BibitemShut {NoStop}%
\bibitem [{\citenamefont {Stixrude}\ \emph {et~al.}(2020)\citenamefont
  {Stixrude}, \citenamefont {Baroni},\ and\ \citenamefont
  {Grasselli}}]{noi2020}%
  \BibitemOpen
  \bibfield  {author} {\bibinfo {author} {\bibfnamefont {L.}~\bibnamefont
  {Stixrude}}, \bibinfo {author} {\bibfnamefont {S.}~\bibnamefont {Baroni}}, \
  and\ \bibinfo {author} {\bibfnamefont {F.}~\bibnamefont {Grasselli}},\
  }\href@noop {} {\enquote {\bibinfo {title} {Thermal evolution of {U}ranus
  with a frozen interior},}\ } (\bibinfo {year} {2020}),\ \Eprint
  {http://arxiv.org/abs/{2004.01756}} {{arXiv}:{2004.01756}
  [{astro-ph.EP}]} \BibitemShut {NoStop}%
\bibitem [{\citenamefont {Stanley}\ and\ \citenamefont
  {Bloxham}(2006)}]{Stanley2006}%
  \BibitemOpen
  \bibfield  {author} {\bibinfo {author} {\bibfnamefont {Sabine}\ \bibnamefont
  {Stanley}}\ and\ \bibinfo {author} {\bibfnamefont {Jeremy}\ \bibnamefont
  {Bloxham}},\ }\bibfield  {title} {\enquote {\bibinfo {title} {{Numerical
  dynamo models of Uranus' and Neptune's magnetic fields}},}\ }\href {\doibase
  10.1016/j.icarus.2006.05.005} {\bibfield  {journal} {\bibinfo  {journal}
  {Icarus}\ }\textbf {\bibinfo {volume} {184}},\ \bibinfo {pages} {556--572}
  (\bibinfo {year} {2006})}\BibitemShut {NoStop}%
\bibitem [{\citenamefont {Vanderbilt}(1985)}]{Vanderbilt85}%
  \BibitemOpen
  \bibfield  {author} {\bibinfo {author} {\bibfnamefont {David}\ \bibnamefont
  {Vanderbilt}},\ }\bibfield  {title} {\enquote {\bibinfo {title} {Optimally
  smooth norm-conserving pseudopotentials},}\ }\href {\doibase
  10.1103/PhysRevB.32.8412} {\bibfield  {journal} {\bibinfo  {journal} {Phys.
  Rev. B}\ }\textbf {\bibinfo {volume} {32}},\ \bibinfo {pages} {8412--8415}
  (\bibinfo {year} {1985})}\BibitemShut {NoStop}%
\bibitem [{\citenamefont {Schlipf}\ and\ \citenamefont
  {Gygi}(2015)}]{Schlipf2015}%
  \BibitemOpen
  \bibfield  {author} {\bibinfo {author} {\bibfnamefont {M}~\bibnamefont
  {Schlipf}}\ and\ \bibinfo {author} {\bibfnamefont {F}~\bibnamefont {Gygi}},\
  }\bibfield  {title} {\enquote {\bibinfo {title} {Optimization algorithm for
  the generation of oncv pseudopotentials},}\ }\href {\doibase
  https://doi.org/10.1016/j.cpc.2015.05.011} {\bibfield  {journal} {\bibinfo
  {journal} {Computer Physics Communications}\ }\textbf {\bibinfo {volume}
  {196}},\ \bibinfo {pages} {36 -- 44} (\bibinfo {year} {2015})},\ \bibinfo
  {note} {with pseudopotentials downloaded from
  \url{http://www.quantum-simulation.org/potentials/sg15_oncv/upf/}}\BibitemShut
  {NoStop}%
\bibitem [{\citenamefont {Perdew}\ \emph {et~al.}(1996)\citenamefont {Perdew},
  \citenamefont {Burke},\ and\ \citenamefont {Ernzerhof}}]{Perdew1996}%
  \BibitemOpen
  \bibfield  {author} {\bibinfo {author} {\bibfnamefont {John~P.}\ \bibnamefont
  {Perdew}}, \bibinfo {author} {\bibfnamefont {Kieron}\ \bibnamefont {Burke}},
  \ and\ \bibinfo {author} {\bibfnamefont {Matthias}\ \bibnamefont
  {Ernzerhof}},\ }\bibfield  {title} {\enquote {\bibinfo {title} {Generalized
  gradient approximation made simple},}\ }\href {\doibase
  10.1103/PhysRevLett.77.3865} {\bibfield  {journal} {\bibinfo  {journal}
  {Phys. Rev. Lett.}\ }\textbf {\bibinfo {volume} {77}},\ \bibinfo {pages}
  {3865--3868} (\bibinfo {year} {1996})}\BibitemShut {NoStop}%
\bibitem [{\citenamefont {Sun}(2019)}]{SunThesis2019}%
  \BibitemOpen
  \bibfield  {author} {\bibinfo {author} {\bibfnamefont {Jiming}\ \bibnamefont
  {Sun}},\ }\emph {\bibinfo {title} {High Pressure Superionic Ice Phase
  Diagram}},\ \href
  {http://physics.princeton.edu/archives/theses/lib/upload/thesis_JimingSun.pdf}
  {Ph.D. thesis},\ \bibinfo  {school} {Princeton University} (\bibinfo {year}
  {2019})\BibitemShut {NoStop}%
\bibitem [{\citenamefont {Nos{\'e}}(1984)}]{Nose1984}%
  \BibitemOpen
  \bibfield  {author} {\bibinfo {author} {\bibfnamefont {Shuichi}\ \bibnamefont
  {Nos{\'e}}},\ }\bibfield  {title} {\enquote {\bibinfo {title} {A unified
  formulation of the constant temperature molecular dynamics methods},}\
  }\href@noop {} {\bibfield  {journal} {\bibinfo  {journal} {The Journal of
  chemical physics}\ }\textbf {\bibinfo {volume} {81}},\ \bibinfo {pages}
  {511--519} (\bibinfo {year} {1984})}\BibitemShut {NoStop}%
\bibitem [{\citenamefont {Hoover}(1985)}]{Hoover1985}%
  \BibitemOpen
  \bibfield  {author} {\bibinfo {author} {\bibfnamefont {William~G}\
  \bibnamefont {Hoover}},\ }\bibfield  {title} {\enquote {\bibinfo {title}
  {Canonical dynamics: Equilibrium phase-space distributions},}\ }\href@noop {}
  {\bibfield  {journal} {\bibinfo  {journal} {Physical review A}\ }\textbf
  {\bibinfo {volume} {31}},\ \bibinfo {pages} {1695} (\bibinfo {year}
  {1985})}\BibitemShut {NoStop}%
\bibitem [{\citenamefont {Martyna}\ \emph {et~al.}(1992)\citenamefont
  {Martyna}, \citenamefont {Klein},\ and\ \citenamefont
  {Tuckerman}}]{Martyna:1992gy}%
  \BibitemOpen
  \bibfield  {author} {\bibinfo {author} {\bibfnamefont {Glenn~J}\ \bibnamefont
  {Martyna}}, \bibinfo {author} {\bibfnamefont {Michael~L}\ \bibnamefont
  {Klein}}, \ and\ \bibinfo {author} {\bibfnamefont {Mark}\ \bibnamefont
  {Tuckerman}},\ }\bibfield  {title} {\enquote {\bibinfo {title}
  {{Nos\'e-Hoover chains: The canonical ensemble via continuous dynamics}},}\
  }\href {\doibase dx.doi.org/10.1063/1.463940} {\bibfield  {journal} {\bibinfo
   {journal} {J. Chem. Phys.}\ }\textbf {\bibinfo {volume} {97}},\ \bibinfo
  {pages} {2635--2643} (\bibinfo {year} {1992})}\BibitemShut {NoStop}%
\bibitem [{\citenamefont {Parrinello}\ and\ \citenamefont
  {Rahman}(1980)}]{Parrinello1980}%
  \BibitemOpen
  \bibfield  {author} {\bibinfo {author} {\bibfnamefont {M\_}\ \bibnamefont
  {Parrinello}}\ and\ \bibinfo {author} {\bibfnamefont {A}~\bibnamefont
  {Rahman}},\ }\bibfield  {title} {\enquote {\bibinfo {title} {Crystal
  structure and pair potentials: A molecular-dynamics study},}\ }\href@noop {}
  {\bibfield  {journal} {\bibinfo  {journal} {Physical Review Letters}\
  }\textbf {\bibinfo {volume} {45}},\ \bibinfo {pages} {1196} (\bibinfo {year}
  {1980})}\BibitemShut {NoStop}%
\bibitem [{\citenamefont {Parrinello}\ and\ \citenamefont
  {Rahman}(1981)}]{Parrinello1981}%
  \BibitemOpen
  \bibfield  {author} {\bibinfo {author} {\bibfnamefont {Michele}\ \bibnamefont
  {Parrinello}}\ and\ \bibinfo {author} {\bibfnamefont {Aneesur}\ \bibnamefont
  {Rahman}},\ }\bibfield  {title} {\enquote {\bibinfo {title} {Polymorphic
  transitions in single crystals: A new molecular dynamics method},}\
  }\href@noop {} {\bibfield  {journal} {\bibinfo  {journal} {Journal of Applied
  physics}\ }\textbf {\bibinfo {volume} {52}},\ \bibinfo {pages} {7182--7190}
  (\bibinfo {year} {1981})}\BibitemShut {NoStop}%
\bibitem [{\citenamefont {Bernasconi}\ \emph {et~al.}(1995)\citenamefont
  {Bernasconi}, \citenamefont {Chiarotti}, \citenamefont {Focher},
  \citenamefont {Scandolo}, \citenamefont {Tosatti},\ and\ \citenamefont
  {Parrinello}}]{Bernasconi1995}%
  \BibitemOpen
  \bibfield  {author} {\bibinfo {author} {\bibfnamefont {M.}~\bibnamefont
  {Bernasconi}}, \bibinfo {author} {\bibfnamefont {G.L.}\ \bibnamefont
  {Chiarotti}}, \bibinfo {author} {\bibfnamefont {P.}~\bibnamefont {Focher}},
  \bibinfo {author} {\bibfnamefont {S.}~\bibnamefont {Scandolo}}, \bibinfo
  {author} {\bibfnamefont {E.}~\bibnamefont {Tosatti}}, \ and\ \bibinfo
  {author} {\bibfnamefont {M.}~\bibnamefont {Parrinello}},\ }\bibfield  {title}
  {\enquote {\bibinfo {title} {First-principle-constant pressure molecular
  dynamics},}\ }\href {\doibase 10.1016/0022-3697(94)00228-2} {\bibfield
  {journal} {\bibinfo  {journal} {Journal of Physics and Chemistry of Solids}\
  }\textbf {\bibinfo {volume} {56}},\ \bibinfo {pages} {501--505} (\bibinfo
  {year} {1995})}\BibitemShut {NoStop}%
\bibitem [{\citenamefont {Zhang}\ \emph {et~al.}(2018)\citenamefont {Zhang},
  \citenamefont {Han}, \citenamefont {Wang}, \citenamefont {Car},\ and\
  \citenamefont {E}}]{Zhang2018}%
  \BibitemOpen
  \bibfield  {author} {\bibinfo {author} {\bibfnamefont {Linfeng}\ \bibnamefont
  {Zhang}}, \bibinfo {author} {\bibfnamefont {Jiequn}\ \bibnamefont {Han}},
  \bibinfo {author} {\bibfnamefont {Han}\ \bibnamefont {Wang}}, \bibinfo
  {author} {\bibfnamefont {Roberto}\ \bibnamefont {Car}}, \ and\ \bibinfo
  {author} {\bibfnamefont {Weinan}\ \bibnamefont {E}},\ }\bibfield  {title}
  {\enquote {\bibinfo {title} {Deep potential molecular dynamics: A scalable
  model with the accuracy of quantum mechanics},}\ }\href {\doibase
  10.1103/PhysRevLett.120.143001} {\bibfield  {journal} {\bibinfo  {journal}
  {Phys. Rev. Lett.}\ }\textbf {\bibinfo {volume} {120}},\ \bibinfo {pages}
  {143001} (\bibinfo {year} {2018})}\BibitemShut {NoStop}%
\bibitem [{\citenamefont {Puligheddu}\ and\ \citenamefont
  {Galli}(2020)}]{Puligheddu2020}%
  \BibitemOpen
  \bibfield  {author} {\bibinfo {author} {\bibfnamefont {Marcello}\
  \bibnamefont {Puligheddu}}\ and\ \bibinfo {author} {\bibfnamefont {Giulia}\
  \bibnamefont {Galli}},\ }\bibfield  {title} {\enquote {\bibinfo {title}
  {Atomistic simulations of the thermal conductivity of liquids},}\ }\href
  {\doibase 10.1103/PhysRevMaterials.4.053801} {\bibfield  {journal} {\bibinfo
  {journal} {Phys. Rev. Materials}\ }\textbf {\bibinfo {volume} {4}},\ \bibinfo
  {pages} {053801} (\bibinfo {year} {2020})}\BibitemShut {NoStop}%
\bibitem [{\citenamefont {Baroni}\ \emph {et~al.}(2001)\citenamefont {Baroni},
  \citenamefont {de~Gironcoli}, \citenamefont {{Dal Corso}},\ and\
  \citenamefont {Giannozzi}}]{Baroni2001}%
  \BibitemOpen
  \bibfield  {author} {\bibinfo {author} {\bibfnamefont {S}~\bibnamefont
  {Baroni}}, \bibinfo {author} {\bibfnamefont {S}~\bibnamefont {de~Gironcoli}},
  \bibinfo {author} {\bibfnamefont {A}~\bibnamefont {{Dal Corso}}}, \ and\
  \bibinfo {author} {\bibfnamefont {P}~\bibnamefont {Giannozzi}},\ }\bibfield
  {title} {\enquote {\bibinfo {title} {Phonons and related crystal properties
  from density-functional perturbation theory},}\ }\href {\doibase
  10.1103/RevModPhys.73.515} {\bibfield  {journal} {\bibinfo  {journal} {Rev.
  Mod. Phys.}\ }\textbf {\bibinfo {volume} {73}},\ \bibinfo {pages} {515--562}
  (\bibinfo {year} {2001})}\BibitemShut {NoStop}%
\bibitem [{\citenamefont {Ercole}\ \emph {et~al.}(2017)\citenamefont {Ercole},
  \citenamefont {Marcolongo},\ and\ \citenamefont {Baroni}}]{Ercole2017}%
  \BibitemOpen
  \bibfield  {author} {\bibinfo {author} {\bibfnamefont {Loris}\ \bibnamefont
  {Ercole}}, \bibinfo {author} {\bibfnamefont {Aris}\ \bibnamefont
  {Marcolongo}}, \ and\ \bibinfo {author} {\bibfnamefont {Stefano}\
  \bibnamefont {Baroni}},\ }\bibfield  {title} {\enquote {\bibinfo {title}
  {Accurate thermal conductivities from optimally short molecular dynamics
  simulations},}\ }\href {\doibase 10.1038/s41598-017-15843-2} {\bibfield
  {journal} {\bibinfo  {journal} {Sci. Rep.}\ }\textbf {\bibinfo {volume}
  {7}},\ \bibinfo {pages} {15835} (\bibinfo {year} {2017})}\BibitemShut
  {NoStop}%
\bibitem [{\citenamefont {Akaike}(1972)}]{Akaike1973}%
  \BibitemOpen
  \bibfield  {author} {\bibinfo {author} {\bibfnamefont {H.}~\bibnamefont
  {Akaike}},\ }\href@noop {} {\emph {\bibinfo {title} {Information theory and
  an extension of the maximum likelihood principle, in 2nd International
  Symposium on Information Theory}}}\ (\bibinfo  {publisher} {edited by B. N.
  Petrov and F. Csáki},\ \bibinfo {year} {1972})\ pp.\ \bibinfo {pages}
  {267--281}\BibitemShut {NoStop}%
\bibitem [{\citenamefont {Akaike}(1974)}]{Akaike1974}%
  \BibitemOpen
  \bibfield  {author} {\bibinfo {author} {\bibfnamefont {H.}~\bibnamefont
  {Akaike}},\ }\bibfield  {title} {\enquote {\bibinfo {title} {A new look at
  the statistical model identification},}\ }\href {\doibase
  10.1109/TAC.1974.1100705} {\bibfield  {journal} {\bibinfo  {journal} {IEEE
  Trans. Autom. Control}\ }\textbf {\bibinfo {volume} {19}},\ \bibinfo {pages}
  {716--723} (\bibinfo {year} {1974})}\BibitemShut {NoStop}%
\bibitem [{\citenamefont {Weisstein}()}]{MovingAverage}%
  \BibitemOpen
  \bibfield  {author} {\bibinfo {author} {\bibfnamefont {Eric~W.}\ \bibnamefont
  {Weisstein}},\ }\href {http://mathworld.wolfram.com/MovingAverage.html}
  {\enquote {\bibinfo {title} {Moving average},}\ }\bibinfo {note} {From
  MathWorld -- a Wolfram Web Resource
  \url{http://mathworld.wolfram.com/MovingAverage.html}}\BibitemShut {NoStop}%
\bibitem [{\citenamefont {Goldsby}\ and\ \citenamefont
  {Kohlstedt}(2001)}]{Goldsby2001}%
  \BibitemOpen
  \bibfield  {author} {\bibinfo {author} {\bibfnamefont {DL}~\bibnamefont
  {Goldsby}}\ and\ \bibinfo {author} {\bibfnamefont {David~L}\ \bibnamefont
  {Kohlstedt}},\ }\bibfield  {title} {\enquote {\bibinfo {title} {Superplastic
  deformation of ice: Experimental observations},}\ }\href@noop {} {\bibfield
  {journal} {\bibinfo  {journal} {Journal of Geophysical Research: Solid
  Earth}\ }\textbf {\bibinfo {volume} {106}},\ \bibinfo {pages} {11017--11030}
  (\bibinfo {year} {2001})}\BibitemShut {NoStop}%
\end{thebibliography}

%

\end{document}


\title{Supplementary Information to \\ ``Heat {and charge} transport in H$_2$O at ice-giant conditions \\ from ab initio molecular dynamics simulations''}

\author{Federico Grasselli}
\altaffiliation{Present affiliation: COSMO -- Laboratory of Computational Science and Modelling, IMX, \'Ecole Polytechnique F\'ed\'erale de Lausanne, 1015 Lausanne, Switzerland}
\affiliation{SISSA -- Scuola Internazionale Superiore di Studi Avanzati, Trieste, Italy}
\author{Lars Stixrude}
\affiliation{Department of Earth, Planetary, and Space Sciences, University of California Los Angeles, USA}
\author{Stefano Baroni}\email{baroni@sissa.it}
\affiliation{SISSA -- Scuola Internazionale Superiore di Studi Avanzati, Trieste, Italy}
\affiliation{CNR -- Istituto Officina dei Materiali, SISSA, 34136 Trieste}

\maketitle

\section{Ab initio simulations}
\textit{Ab initio} molecular dynamics simulations were run using the plane-wave pseudo-potential method and the Car-Parrinello (CP) Lagrangian formalism \cite{Car1985}, as implemented in the {\texttt{cp.x} component of the} \textsc{Quantum ESPRESSO} suite of computer codes \cite{Giannozzi2009,Giannozzi2017}. We adopted norm-conserving  pseudopotentials \cite{Vanderbilt85} from Ref. \onlinecite{Schlipf2015}, which are accurate also at the present pT-conditions \cite{Sun2015}, and the PBE energy functional \cite{Perdew1996}. We employed a plane-wave energy cutoff of $E_\mathrm{cut}= 110\un{Ry}$ for wave-functions and four times as much for the charge density. 
The fictitious electronic mass was set to 25 electronic masses and the integration time step to $\approx 0.0484$ fs. These choices ensure adequate adiabatic decoupling of the electron fictitious dynamics from the ionic motion and conservation of the extended total energy in the relevant pT range considered here.
{All our simulations have been performed neglecting nuclear quantum effects (NQEs), as these are expected to become negligible as temperature increases. Even though NQEs may be counter-enhanced by high pressure \cite{SunThesis2019}, their account is beyond the scope of the present work.}

Flux time series were collected from $NVE$ simulations for at least 60 ps, after an initial $NPT$ equilibration of a few ps at the target pT conditions, followed by $\sim 4 \un{ps}$ of further $NVE$ equilibration.
$NPT$ simulations were run with a chain of three Nos\'e-Hoover thermostats \cite{Nose1984,*Hoover1985,*Martyna:1992gy} with a frequency of $60\un{THz}$, while pressure was controlled via a Parrinello-Rahman barostat \cite{Parrinello1980,*Parrinello1981}. In order to minimize the effects of Pulay stresses, the kinetic energy functional was modified as: $\mathbf{G}^2 \mapsto \mathbf{G}^2 + A \left[ 1 + \mathrm{erf}[ (\mathbf{G}^2- E_0)/\sigma] \right]$ \cite{Bernasconi1995} with $E_0 = 100\un{Ry}$, $A = 170\un{Ry}$, and  $\sigma = 15\un{Ry}$. 

{Our simulations were run on samples of 128 atoms for ice-X, SI-BCC, and PDL water, and of 108 atoms for SI-FCC. In order to estimate the magnitude of finite-size effects arising in the AIMD simulation of the thermal conductivity, we have performed extensive MD simulations of PDL water samples of 128, 256, 512, and 1024 H$_2$O units, using a deep-neural-network potential trained on our ab initio trajectory \cite{Zhang2018}, at the same conditions ($T\approx 1970~\mathrm{K}$ and $p\approx 33~\mathrm{GPa}$). These results indicate that finite-size corrections to the thermal conductivity computed for our sample of 128 H$_2$O units should be smaller than 5\%, in qualitative agreement with systematic tests performed in Ref. \onlinecite{Puligheddu2020}.}



\section{Flux time series}
\begin{figure}
    \centering
    \includegraphics[width=\columnwidth]{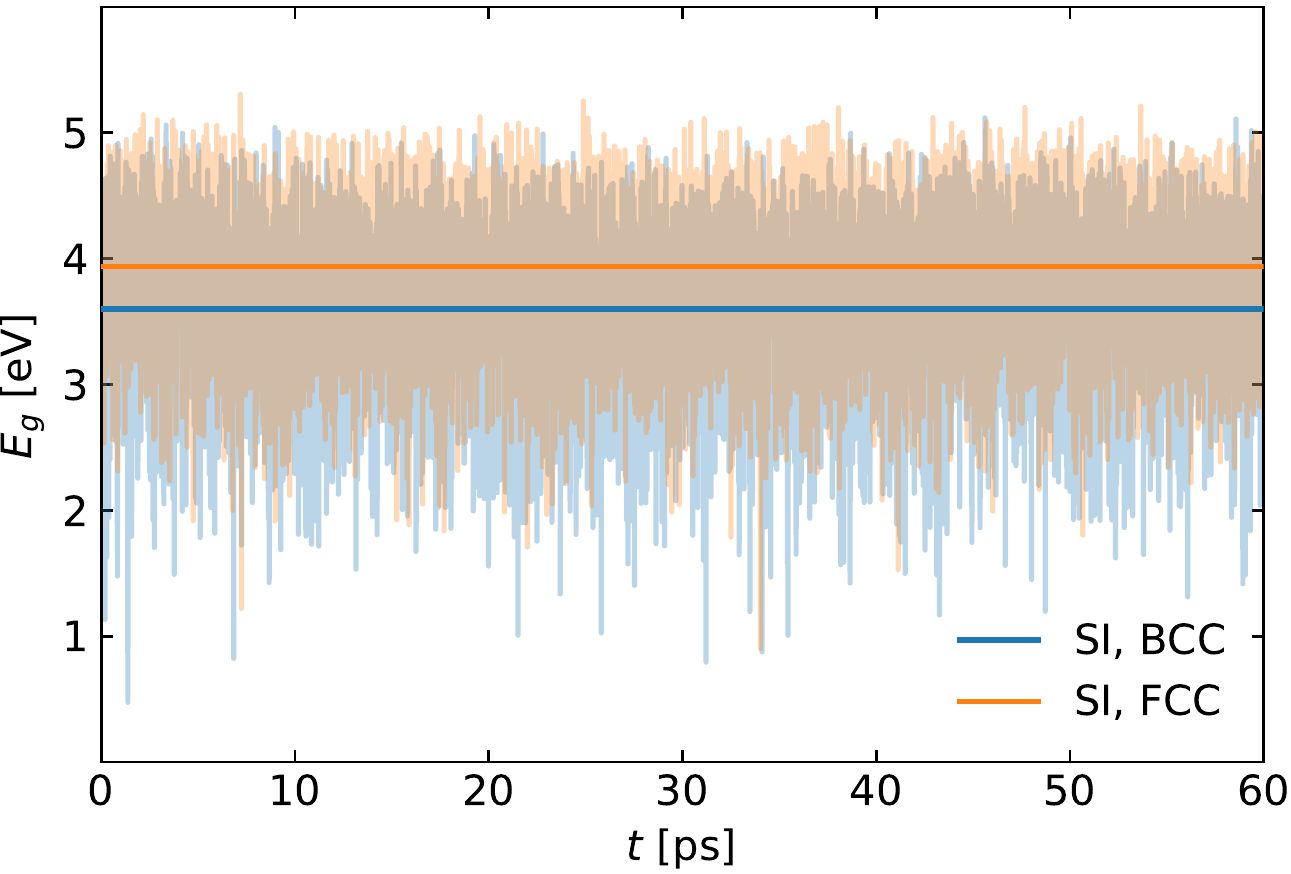}
    \caption{Energy band gap, $E_g$, during the CP AIMD trajectory (transparent), and its average (solid horizontal lines). Values for the superionic phase with oxygen BCC lattice (blue, $2950 \text{ K, } 171 \text{ GPa}$) and FCC lattice (orange, $2920 \text{ K, } 257 \text{ GPa}$). }
    \label{fig:gap}
\end{figure}

The energy and charge fluxes, $\mathcal{J}_E(t)$ and $\mathcal{J}_Z(t)$, were sampled every 30 or 40 time steps of the original CP simulation ($\Delta_t = 1.45\un{fs}$ or $1.94\un{fs}$), depending on the specific simulation. This sampling rate is high enough to make the flux power spectra vanish at the corresponding Nyqvist frequency, $f_\mathrm{Ny} = 1/(2\Delta_t)$, thus avoiding any aliasing effects. We verified that the energy band gap is far larger than $k_BT$ all along our simulations, thus excluding any possible non adiabatic effects in heat and charge transport at the pT conditions of interest here. The time series of the energy gap occurring in two different SI phases of water are reported in Fig. \ref{fig:gap}.

The energy flux was estimated using the expression derived in Ref. \onlinecite{Marcolongo2016}. The charge flux was estimated using two different expressions: \emph{i)} the charge flux, $\mathcal{J}_Z$, estimated from the classical expression derived from the formal oxidation numbers $q_\mathrm{H}=+1$ and $q_\mathrm{O}=-2$, according to the topological theory of adiabatic charge transport introduced in Ref. \onlinecite{Grasselli2019}, and \emph{ii)} the standard quantum-mechanical one, $\mathcal{J}'_Z$, that is the sum of the adiabatic electron-charge flux, $\mathcal{J}_{el}$, and the classical current of the atomic cores ($Z_\mathrm{H}=+1$ and $Z_\mathrm{O}=+6$). The two expressions read:
\begin{align}
    \mathcal{J}_Z &=\frac{e}{\Omega} \left( q_\mathrm{H} \sum_{n\in \mathrm{H}} \mathcal{V}_n + q_\mathrm{O} \sum_{n\in \mathrm{O}} \mathcal{V}_n \right) \label{eq:JZ} \\
    \mathcal{J}'_Z &= \frac{e}{\Omega} \left( Z_\mathrm{H} \sum_{n\in \mathrm{H}} \mathcal{V}_n  
    + Z_\mathrm{O} \sum_{n\in \mathrm{O}} \mathcal{V}_n \right) + \mathcal{J}_{el}. \label{eq:JZ'}
\end{align}
The adiabatic electron charge current, $\mathcal{J}_{el}$, can be computed as:
\begin{equation}
 \mathcal{J}_{el} = \frac{-2e}{\Omega}\mathfrak{Re} \sum_{v} \langle\bm{\bar \phi}_v^c |\dot \phi_v^c \rangle , \label{eq:Jel}
\end{equation}
where the sum runs over the occupied Kohn-Sham states, $\{\phi_v\}$, and the orbitals 
\begin{equation}
    \begin{aligned}
        |\bm{\bar \phi}_v^c\rangle &= \hat P_c \,\mathbf{r}  \,|\phi_v \rangle, \\
        |\dot \phi_v^c\rangle &= \dot{\hat P}_v \,|\phi_v\rangle, 
    \end{aligned}
    \label{eq:phi-dot}
\end{equation}
are the projections over the empty-state manifold of the action of the position operator over the $v$-th occupied orbital, and of its adiabatic time derivative \citep{Giannozzi2017}. The operators $\hat P_v$ and $\hat P_c = 1 - \hat P_v$ are projectors over the occupied- and empty-states manifolds, respectively. Equations \eqref{eq:phi-dot} are well defined in {periodic boundary conditions} and have been computed from standard density-functional perturbation theory \citep{Baroni2001}. 



\section{Cepstral analysis}
Transport coefficients were estimated by cepstral analysis of the flux time series, as explained in Refs. \onlinecite{Ercole2017} and \onlinecite{Bertossa2019}, using the \texttt{thermocepstrum} code \cite{thermocepstrum}. This technique allows one to obtain accurate transport coefficients, as well as a quantitative estimate of their statistical accuracy, depending on two parameters: an effective Nyqvist frequency, $f^*$, used to limit the analysis to a properly defined low-frequency portion of the spectrum, and the number $P^*$ of inverse Fourier (``\emph{cepstral}'') coefficients of the logarithm of the spectrum in the low-frequency region thus defined. The estimated conductivities depend very little on $f^*$, whereas an optimal value of $P^*$ can be estimated in most cases by statistical model-selection techniques, as explained in Ref. \onlinecite{Ercole2017}. In our applications $P^*$ is determined via the so-called \emph{Akaike Information Criterion} \cite{Akaike1973,*Akaike1974,Ercole2017} 

\begin{table*}[]
    \centering
    \begin{tabular}{
    c 
    S[table-format=3] 
    S[table-format=1.2]
    S[table-format=4(2)]
    S[table-format=4(3)]
    S[table-format=2.2]
    S[table-format=2]
    S[table-format=3.1(3)]
    S[table-format=3(2)]
    c}
    \toprule
         \multirow{2}*{phase} 
        & $N_\mathrm{H_2O}$ & $\rho$
        & $T$ & $p$  
        & $f^\ast$ & $\textit{P}^\ast$
        & $\kappa$ & $\sigma$ & $\textit{D}$
         \\
        &  & \si{[\gram\per\centi\meter\tothe{3}]}
        & \si{[\kelvin]} & \si{[\giga\pascal]} 
        & \si{[\tera\hertz]} & 
        & \si{[\watt\per(\kelvin\meter)]} & \si{[\siemens\per\centi\meter]} & \si{[\angstrom\tothe{2}/\pico\second]}
         \\
        \cmidrule{1-10}
        ice X    & 128     & 3.52     & 1488 \pm 45     & 182 \pm 1   & 13.8  & 6   & 16.1 \pm 1.1 &  $\mathrm{-}$ & $\mathrm{consist.\,zero}$\\ [1.0ex]
        SI$^\mathrm{BCC}$ & 128     & 3.39     & 2474 \pm 78     &  174 \pm 2    & 31.3  & 3  &  9.4 \pm 0.6    &  135 \pm 7  &  $4.88\pm 0.13$   (H) \\
        SI$^\mathrm{BCC}$ & 128     &  3.35      &  2945 \pm 88      &  171 \pm 2    & 36.9  & 5   &  10.7 \pm 0.7   &  180 \pm 5  & $7.41 \pm 0.12  $ (H)\\
        SI$^\mathrm{BCC}$ & 128     &  3.61      &  2905 \pm 86      &  218 \pm 2    & 31.3  & 3   &  9.9 \pm 0.7    &  198 \pm 9  & $7.38\pm 0.16  $ (H)\\ 
        SI$^\mathrm{FCC}$ & 108     &  3.82      &  2917 \pm 93      &  257 \pm 2    & 38.3  & 5  &  12.8 \pm 1.0   &  256 \pm 8  & $7.17 \pm 0.13 $ (H)\\ [1.0ex]
        PDL   & 128   &  2.04      &  1970 \pm 60      &  33 \pm 1   & 49.2  & 5   &  4.1 \pm 0.3    &  42 \pm 3  & $3.10 \pm 0.03$ (H) \\
        & & & & & & & & & $0.92\pm 0.02$ (O) \\ [1.0ex]
        d-ice X     & 127  &  3.49      &  1523 \pm 47   &  179 \pm 1  & 18.1 & 8 &  17.0 \pm 1.5   &  $\mathrm{<0.1}$  & $0.0039 \pm 0.0006$  (H)\\ [1.0ex]
        d-SI$^\mathrm{BCC}$   & 127 &  3.49      &  2302 \pm 94      &  190 \pm 2     & 38.3 & 6 &  11.6 \pm 0.9   &  95 \pm 3  & $3.04 \pm 0.04$ (H) \\
        & & & & & & & & & $0.0085 \pm 0.0006$  (O)  \\
        \bottomrule
    \end{tabular}
    \caption{Results of the different NVE simulations. $N_\mathrm{H_2O}$ is the number of H$_2$O stoichiometric units employed in the simulation. Temperature and pressures are reported together with their standard deviation. The last two simulations concern the defected structure, where one H$_2$O was removed. The temperatures and pressures are reported with their standard deviations. The uncertainty on the diffusivity is estimated via a 6 block analysis with $\approx 10$-ps segments.}
    \label{tab:NVE_results}
\end{table*}

\begin{figure}
    \centering
    \includegraphics[width=\columnwidth]{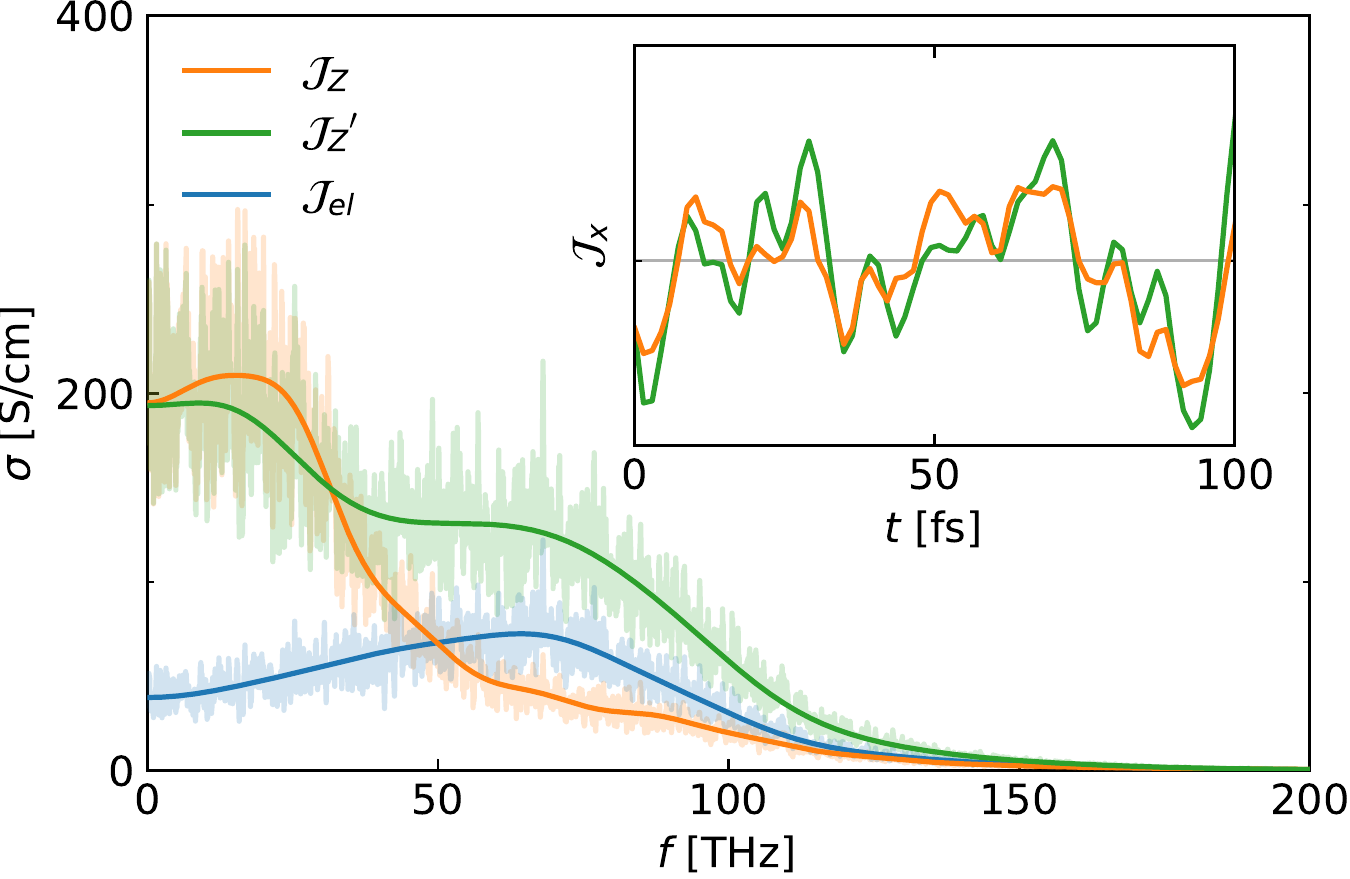}
    \caption{Periodograms (filtered with a 0.1 THz window) and cepstral analysis of electric fluxes, as defined in Eqs. \eqref{eq:JZ}, \eqref{eq:JZ'} and \eqref{eq:Jel} and following, for the simulation of the SI-BCC phase at $T=2910 \un{K}$ and $P=218 \un{GPa}$. Inset: a sample of the time series of the the electric fluxes ($x-$component), whose zero is highlighted by the horizontal gray line.}
    \label{fig:elcurr}
\end{figure}

Figure \ref{fig:elcurr} displays the sample power spectra of $\mathcal{J}_{el}$, $\mathcal{J}_Z$ and $\mathcal{J}'_Z$, scaled in units such that the zero-frequency value is the electrical conductivity in S/cm. The faint lines are obtained by applying a moving average filter \cite{MovingAverage} with a window of 0.1 THz to the raw periodograms; the solid smooth lines are obtained by applying a cepstral filter \cite{Ercole2017} with the parameters reported in Table \ref{tab:NVE_results}. The definitions $\mathcal{J}_Z$ and $\mathcal{J}'_Z$, lead to the same conductivity, even if their time series values differ, as is clear from the inset, in agreement with Ref. \cite{Grasselli2019}, where this finding was proved to be the consequence of gauge-invariance of transport coefficients and topological quantization of adiabatic charge transport in electronically gapped materials. 


\begin{figure}
    \centering
    \includegraphics[width=\columnwidth]{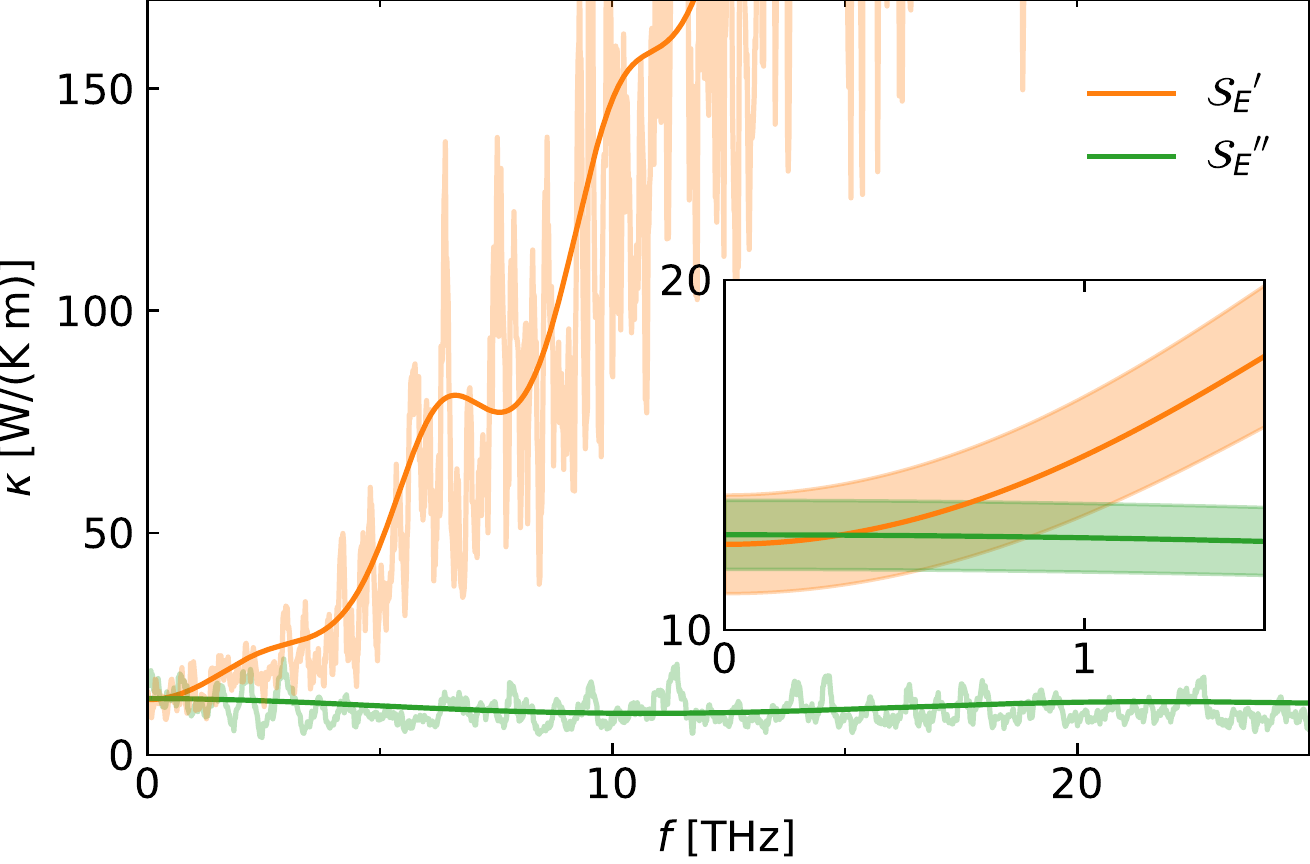}
    \caption{Energy-flux power spectra for FCC super-ionic water at average $T = 2920\un{K}$ and $P = 257\un{GPa}$. Orange: 2-variate reduced periodogram, $\mathcal{S}'_E$. 
   {Green: 3-variate reduced periodogram obtained under the constraint of vanishing mass and adiabatic electronic fluxes, $\mathcal{S}''_E$. The reported data are the result of a 0.2 THz window filtering.} 
    The smooth thick lines are the filtered spectra obtained via 
    cepstral analysis. Inset: low-frequency zoom of with their estimated uncertainties.}
    \label{fig:FCC_SM}
\end{figure}

As stated in the main text, a 3-variate analysis on thermal conductivity adopting the adiabatic electron flux $\mathcal{J}_{el}$ as an additional flux reduces the statistical uncertainty without affecting the estimate of the thermal conductivity, as shown in Fig. \ref{fig:FCC_SM} (green). The 3-variate analysis produces a flatter spectrum featuring a reduced total {power}, corresponding to a smaller number of cepstral coefficients, $P^\ast$, thus ensuring a smaller statistical error. 

A comprehensive summary of our results, along with the values of $f^*$ and $P^*$ used for cepstral analysis, is given in Tab. \ref{tab:NVE_results}. 

\section{Defected structure}
\begin{figure}
    \centering
    \includegraphics[width=\columnwidth]{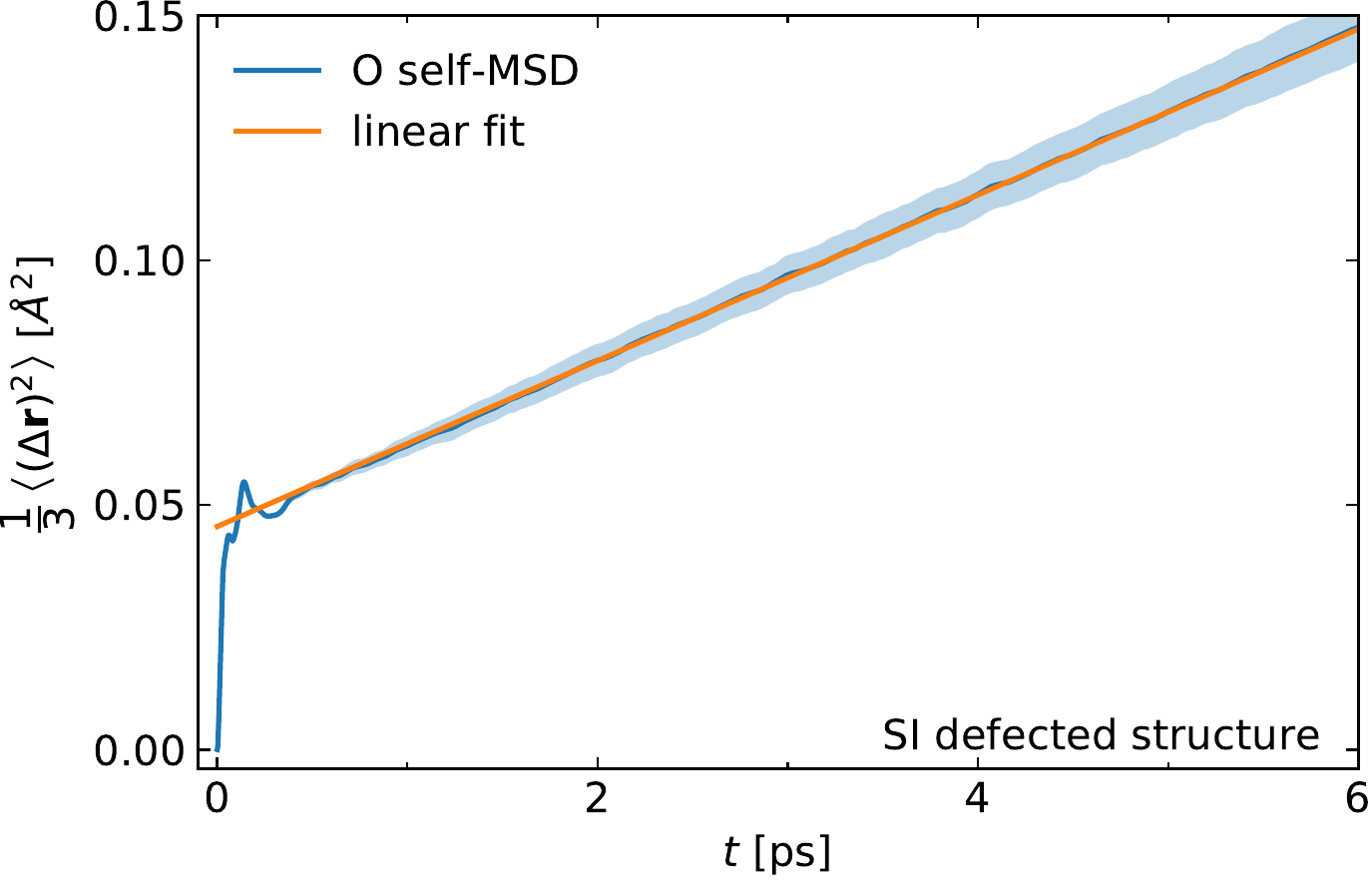}
    \caption{Mean square displacement of oxygen atoms (blue solid) and its statistical uncertainty from a 6 block analysis of 10ps segments.}
    \label{fig:MSD-O-def}
\end{figure}

We also performed two further simulations---one in solid ice X and one in the SI phase with BCC oxygen structure---for defected structures where two H and one O atoms were removed, to preserve the stoichiometry of the ideal systems, resulting in a defect relative concentration of $1/128\approx 0.008$.
In the SI phase, vacancy hopping in the oxygen lattice is observed in the form of jumps of one or more than one O atoms within the same hopping event. From the mean square displacement of oxygen atoms with respect to their center of mass, we estimated the O self-diffusion coefficient as $D_\mathrm{O} \approx (8.5 \pm 0.6)\times 10^{-3} \un{\AA^2/ps}$. The linear growth of the oxygen mean square displacement, together with the related statistical uncertainty from a  6-block analysis of 10ps segments, is reported in Fig. \ref{fig:MSD-O-def}.
On the contrary, in the defected ice X structure below the SI transition we observed no oxygen self-diffusion.

Even if such defect concentration is far larger than what it is expected to be at the pT-conditions of Uranus and Neptune from extrapolation of experimental data \cite{Goldsby2001}, it is still insufficient to strongly affect the thermal conductivity, when compared to the ideal-structure value. In fact, we obtained values for $\kappa$ which are compatible with those of the ideal structures, namely $\kappa \approx 17$ and $\kappa \approx 12$ \si{\watt\per(\kelvin\meter)} for the solid and the SI phases, respectively.

%